\documentclass[a4paper]{IEEEtran}
\usepackage{amsfonts,amssymb}
\usepackage[utf8]{inputenc} 
\usepackage[T1]{fontenc}
\usepackage{url}
\usepackage{ifthen}
\usepackage{cite}
\usepackage[cmex10]{amsmath}
\newtheorem{thm}{Theorem}
\newtheorem{cor}[thm]{Corollary}
\newtheorem{lem}{Lemma}
\newtheorem{df}{Definition}
\newtheorem{rem}{Remark}

\newcount\pictmp
\def\ave#1#2#3{
 \pictmp=#2
 \advance\pictmp by #3
 \ifodd\pictmp
 \divide\pictmp by 2
 \edef#1{\number\pictmp.5}
 \else
 \divide\pictmp by 2
 \edef#1{\number\pictmp}
 \fi
}
\def\qbline(#1,#2)(#3,#4){
 \ave{\picx}{#1}{#3}\ave{\picy}{#2}{#4}
 \qbezier(#1,#2)(\picx,\picy)(#3,#4)
}

\newcommand{\fC}{\mathfrak{C}}
\newcommand{\NN}{\mathbb{N}}
\newcommand{\C}{\mathcal{C}}
\newcommand{\D}{\mathcal{D}}
\newcommand{\F}{\mathcal{F}}
\newcommand{\I}{\mathcal{I}}
\newcommand{\J}{\mathcal{J}}
\newcommand{\M}{\mathcal{M}}
\newcommand{\RITs}{\mathcal{R}_{\mathrm{IT}}^{\mathrm{source}}}
\newcommand{\ROPs}{\mathcal{R}_{\mathrm{OP}}^{\mathrm{source}}}
\newcommand{\RIT}{\mathcal{R}_{\mathrm{IT}}^{\mathrm{channel}}}
\newcommand{\RITD}{\mathcal{R}_{\mathrm{ITD}}^{\mathrm{channel}}}
\newcommand{\RITS}{\mathcal{R}_{\mathrm{ITS}}^{\mathrm{channel}}}
\newcommand{\ROP}{\mathcal{R}_{\mathrm{OP}}^{\mathrm{channel}}}
\newcommand{\cS}{\mathcal{S}}
\newcommand{\T}{\mathcal{T}}
\newcommand{\U}{\mathcal{U}}
\newcommand{\V}{\mathcal{V}}
\newcommand{\X}{\mathcal{X}}
\newcommand{\Y}{\mathcal{Y}}
\newcommand{\Z}{\mathcal{Z}}
\newcommand{\G}{\mathcal{G}}
\newcommand{\zero}{\boldsymbol{0}}
\newcommand{\one}{\boldsymbol{1}}
\newcommand{\aalpha}{\boldsymbol{\alpha}}
\newcommand{\bbeta}{\boldsymbol{\beta}}
\newcommand{\cc}{\boldsymbol{c}}
\newcommand{\mm}{\boldsymbol{m}}
\newcommand{\bp}{\boldsymbol{p}}
\newcommand{\xx}{\boldsymbol{x}}
\newcommand{\yy}{\boldsymbol{y}}
\newcommand{\zz}{\boldsymbol{z}}
\newcommand{\CC}{\boldsymbol{C}}
\newcommand{\MM}{\boldsymbol{M}}
\newcommand{\UU}{\boldsymbol{U}}
\newcommand{\VV}{\boldsymbol{V}}
\newcommand{\WW}{\boldsymbol{W}}
\newcommand{\XX}{\boldsymbol{X}}
\newcommand{\YY}{\boldsymbol{Y}}
\newcommand{\ZZ}{\boldsymbol{Z}}
\newcommand{\bcF}{\boldsymbol{\mathcal{F}}}
\newcommand{\bcG}{\boldsymbol{\mathcal{G}}}
\newcommand{\hmm}{\widehat{\boldsymbol{m}}}
\newcommand{\hzz}{\widehat{\boldsymbol{z}}}
\newcommand{\hM}{\widehat{M}}
\newcommand{\hU}{\widehat{U}}
\newcommand{\hZ}{\widehat{Z}}
\newcommand{\hu}{\widehat{u}}
\newcommand{\tZ}{\widetilde{Z}}
\newcommand{\oO}{\overline{O}}
\newcommand{\bQ}{\overline{Q}}
\newcommand{\e}{\varepsilon}
\newcommand{\Prod}{\operatornamewithlimits{\text{\Large $\times$}}}
\newcommand{\limn}{\lim_{n\to\infty}}
\newcommand{\liminfn}{\liminf_{n\to\infty}}
\newcommand{\limsupn}{\limsup_{n\to\infty}}
\newcommand{\pliminfn}{\operatornamewithlimits{\text{p-liminf}}_{n\to\infty}}
\newcommand{\plimsupn}{\operatornamewithlimits{\text{p-limsup}}_{n\to\infty}}
\newcommand{\Encoder}{\Phi}
\newcommand{\Decoder}{\Psi}
\newcommand{\oH}{\overline{H}}
\newcommand{\uH}{\underline{H}}
\newcommand{\uI}{\underline{I}}
\newcommand{\oT}{\overline{\mathcal{T}}}
\newcommand{\uT}{\underline{\mathcal{T}}}
\newcommand{\markov}{\leftrightarrow}
\newcommand{\im}{\mathrm{Im}}
\newcommand{\lrB}[1]{\left[{#1}\right]}
\newcommand{\lrb}[1]{\left\{{#1}\right\}}
\newcommand{\lrsb}[1]{\left({#1}\right)}
\newcommand{\lrbar}[1]{\left|{#1}\right|}
\newcommand{\Prob}{\mathrm{P}}
\newcommand{\Error}{\mathrm{Error}}

\allowdisplaybreaks

\title{
 Multi-Terminal Codes\\
 Using Constrained-Random-Number Generators
}
\author{
  Jun~Muramatsu
  and~Shigeki Miyake
  \thanks{J.~Muramatsu is with
   NTT Communication Science Laboratories, NTT Corporation,
   2-4, Hikaridai, Seika-cho, Soraku-gun, Kyoto 619-0237, Japan
   (E-mail: muramatsu.jun@lab.ntt.co.jp).
   S.~Miyake is with
   NTT Network Innovation Laboratories, NTT Corporation,
   Hikarinooka 1-1, Yokosuka-shi, Kanagawa 239-0847, Japan
   (E-mail: miyake.shigeki@lab.ntt.co.jp).
  }
}
\begin{document}
\maketitle

\renewcommand{\baselinestretch}{0.93}

\begin{abstract}
 A general multi-terminal source code and a general
 multi-terminal channel code are presented.
 Constrained-random-number generators with sparse matrices,
 which are building blocks for the code construction,
 are used in the construction of both encoders and decoders.
 Achievable regions for source coding and channel coding
 are derived in terms of entropy functions,
 where the capacity region for channel coding 
 provides an alternative to the region of
 [Somekh-Baruch and Verd\'u, ISIT2006].
\end{abstract}

\section{Introduction}

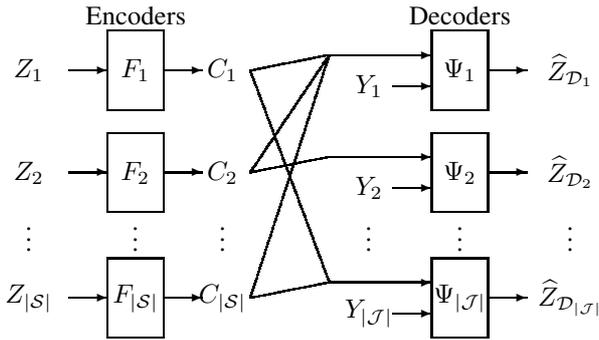
\begin{figure}[h]
\begin{center}
 \unitlength 0.52mm
 \begin{picture}(147,85)(-5,10)
  \put(0,78){\makebox(0,0){$Z_1$}}
  \put(0,52){\makebox(0,0){$Z_2$}}
  \put(0,37){\makebox(0,0){$\vdots$}}
  \put(0,20){\makebox(0,0){$Z_{|\cS|}$}}
  \put(10,78){\vector(1,0){10}}
  \put(10,52){\vector(1,0){10}}
  \put(10,20){\vector(1,0){10}}
  \put(27,92){\makebox(0,0){Encoders}}
  \put(20,68){\framebox(14,20){$F_1$}}
  \put(20,42){\framebox(14,20){$F_2$}}
  \put(27,37){\makebox(0,0){$\vdots$}}
  \put(20,10){\framebox(14,20){$F_{|\cS|}$}}
  \put(34,78){\vector(1,0){10}}
  \put(49,78){\makebox(0,0){$C_1$}}
  \put(34,52){\vector(1,0){10}}
  \put(49,52){\makebox(0,0){$C_2$}}
  \put(49,37){\makebox(0,0){$\vdots$}}
  \put(34,20){\vector(1,0){10}}
  \put(49,20){\makebox(0,0){$C_{|\cS|}$}}
  
  \qbline(56,78)(76,82)
  \qbline(56,78)(76,24)
  \qbline(56,52)(76,82)
  \qbline(56,52)(76,56)
  \qbline(56,20)(76,24)
  \qbline(56,20)(76,82)
  
  \put(109,92){\makebox(0,0){Decoders}}
  \put(76,82){\vector(1,0){26}}
  \put(86,74){\makebox(0,0){$Y_1$}}
  \put(92,74){\vector(1,0){10}}
  \put(102,68){\framebox(14,20){$\Decoder_1$}}
  \put(76,56){\vector(1,0){26}}
  \put(86,48){\makebox(0,0){$Y_2$}}
  \put(92,48){\vector(1,0){10}}
  \put(102,42){\framebox(14,20){$\Decoder_2$}}
  \put(86,37){\makebox(0,0){$\vdots$}}
  \put(109,37){\makebox(0,0){$\vdots$}}
  \put(76,24){\vector(1,0){26}}
  \put(86,16){\makebox(0,0){$Y_{|\J|}$}}
  \put(92,16){\vector(1,0){10}}
  \put(102,10){\framebox(14,20){$\Decoder_{|\J|}$}}
  \put(116,78){\vector(1,0){10}}
  \put(137,78){\makebox(0,0){$\hZ_{\D_1}$}}
  \put(116,52){\vector(1,0){10}}
  \put(137,52){\makebox(0,0){$\hZ_{\D_2}$}}
  \put(137,37){\makebox(0,0){$\vdots$}}
  \put(116,20){\vector(1,0){10}}
  \put(137,20){\makebox(0,0){$\hZ_{\D_{|\J|}}$}}
 \end{picture}
\end{center}
\caption{Multi-terminal Source Coding}
\label{fig:source}
\end{figure}

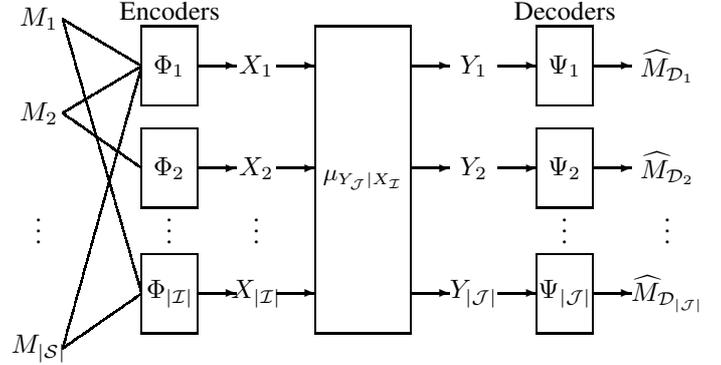
\begin{figure}[h]
\begin{center}
 \unitlength 0.52mm
 \begin{picture}(157,85)(-5,10)
  \put(-6,90){\makebox(0,0){$M_1$}}
  \put(-6,66){\makebox(0,0){$M_2$}}
  \put(-6,38){\makebox(0,0){$\vdots$}}
  \put(-6,6){\makebox(0,0){$M_{|\cS|}$}}
  \qbline(0,90)(20,78)
  \qbline(0,90)(20,20)
  \qbline(0,66)(20,78)
  \qbline(0,66)(20,52)
  \qbline(0,6)(20,20)
  \qbline(0,6)(20,78)
  \put(27,92){\makebox(0,0){Encoders}}
  \put(20,68){\framebox(14,20){$\Encoder_1$}}
  \put(20,42){\framebox(14,20){$\Encoder_2$}}
  \put(27,38){\makebox(0,0){$\vdots$}}
  \put(20,10){\framebox(14,20){$\Encoder_{|\I|}$}}
  \put(34,78){\vector(1,0){10}}
  \put(49,78){\makebox(0,0){$X_1$}}
  \put(54,78){\vector(1,0){10}}
  \put(34,52){\vector(1,0){10}}
  \put(49,52){\makebox(0,0){$X_2$}}
  \put(54,52){\vector(1,0){10}}
  \put(49,38){\makebox(0,0){$\vdots$}}
  \put(34,20){\vector(1,0){10}}
  \put(49,20){\makebox(0,0){$X_{|\I|}$}}
  \put(54,20){\vector(1,0){10}}
  \put(64,10){\framebox(24,78){\small $\mu_{Y_{\J}|X_{\I}}$}}
  \put(127,92){\makebox(0,0){Decoders}}
  \put(88,78){\vector(1,0){10}}
  \put(104,78){\makebox(0,0){$Y_1$}}
  \put(110,78){\vector(1,0){10}}
  \put(120,68){\framebox(14,20){$\Decoder_1$}}
  \put(88,52){\vector(1,0){10}}
  \put(104,52){\makebox(0,0){$Y_2$}}
  \put(110,52){\vector(1,0){10}}
  \put(120,42){\framebox(14,20){$\Decoder_2$}}
  \put(127,38){\makebox(0,0){$\vdots$}}
  \put(88,20){\vector(1,0){10}}
  \put(104,20){\makebox(0,0){$Y_{|\J|}$}}
  \put(110,20){\vector(1,0){10}}
  \put(120,10){\framebox(14,20){$\Decoder_{|\J|}$}}
  \put(134,78){\vector(1,0){10}}
  \put(153,78){\makebox(0,0){$\hM_{\D_1}$}}
  \put(134,52){\vector(1,0){10}}
  \put(153,52){\makebox(0,0){$\hM_{\D_2}$}}
  \put(153,38){\makebox(0,0){$\vdots$}}
  \put(134,20){\vector(1,0){10}}
  \put(153,20){\makebox(0,0){$\hM_{\D_{|\J|}}$}}
 \end{picture}
\end{center}
\caption{Multi-terminal Channel Coding}
\label{fig:channel}
\end{figure}

In this paper, we consider the problems of multi-terminal
source coding (Fig.~\ref{fig:source}) and channel coding
(Fig.~\ref{fig:channel}).
First, we construct a code for correlated sources
and derive an achievable region.
Our setting extends separate coding for correlated
sources~\cite{C75}\cite{MK95}\cite{SW73}.
Next, we use the source code
to construct a code for a general single-hop multi-terminal channel,
which includes multiple-access channels~\cite{H79}\cite{H98}\cite{SW73MAC}
and broadcast channels~\cite{GP80}\cite{IO05}\cite{M79}.
We derive multi-letter characterized capacity regions for these problems
by showing that they are achievable with the constructed codes.
Our capacity region for the channel coding
is specified in terms of entropy functions
and provides an alternative to the region derived in~\cite{SV06}.
It should be noted that,
when auxiliary random variables are assumed to be
independent and identically distributed (i.i.d.),
our region provides the best known single-letter characterized
achievable regions for i.i.d. channels.

Throughout this paper, we use the information spectrum method
introduced in \cite{HAN}, and we do not assume such conditions as consistency,
stationarity and ergodicity.
Let $\Prob(\cdot)$ denote the probability of an event.
For a sequence $\{\mu_{U^n}\}_{n=1}^{\infty}$
of probability distributions corresponding to $\UU\equiv\{U^n\}_{n=1}^{\infty}$,
$\uH(\UU)$ denotes the spectral inf-entropy rate.
For a sequence $\{\mu_{U^nV^n}\}_{n=1}^{\infty}$ of
joint probability distributions corresponding to
$(\UU,\VV)\equiv\{(U^n,V^n)\}_{n=1}^{\infty}$,
$\oH(\UU|\VV)$ denotes the spectral conditional sup-entropy rate.
Formal definitions are given in  Appendix~\ref{sec:ispec}.

We define $\chi(\mathrm{S})\equiv 1$ if $\mathrm{S}$ is true.
Otherwise, we define $\chi(\mathrm{S})\equiv 0$.
The set $\U\setminus\V$ denotes the set difference
and $\U'^c\equiv\U\setminus\U'$.
For a set $f_{\cS}\equiv\{f_s\}_{s\in\cS}$
of functions and a set $\cc_{\cS}\equiv\{\cc_s\}_{s\in\cS}$ of vectors,
let
\begin{align*}
 \fC_{f_{\cS}}(\cc_{\cS})
 &\equiv
 \{\zz_{\cS}: f_s(\zz_s)=\cc_s\ \text{for all}\ s\in\cS\},
\end{align*}
where $\zz_{\cS}\equiv\{\zz_s\}_{s\in\cS}$.

\section{Construction of Source Code}
\label{sec:source}

We introduce the single-hop multi-terminal source coding problem illustrated in
Fig.~\ref{fig:source}.
This code is used to construct a channel code.

For an index set $\cS$ of messages and an index set $\J$ of decoders,
let
$(\ZZ_{\cS},\YY_{\J})\equiv\{(\{Z^n_s\}_{s\in\cS},\{Y^n_j\}_{j\in\J})\}_{n=1}^{\infty}$
be a general correlated source, which is characterized by
joint distributions $\mu_{Z_{\cS}^nY_{\J}^n}$ of
$Z_{\cS}^n,\equiv\{Z_s^n\}_{s\in\cS}$ and
$Y_{\J}^n\equiv\{Y_j^n\}_{j\in\J}$.
For each $s\in\cS$ and $n\in\NN$,
let $\Z_s^n$ be the alphabet of a message $Z_s^n$.
For each $j\in\J$ and $n\in\NN$,
let $\Y_j^n$ be the alphabet of side information $Y_j^n$
available for the $j$-th decoder.
It should be noted that we assume that
$\Z_s^n$ is a finite set
but $\Y^n$ is allowed to be an arbitrary (infinite, continuous) set.

For each $s\in\cS$ and $n\in\NN$,
let $F_{s,n}:\Z_s^n\to\C_{s,n}$ be the $s$-th (possibly stochastic)
encoding function, where
$\C_{s,n}$ is the set of all codewords.
Let $C_{s,n}\equiv F_{s,n}(Z_s^n)$ be the codeword of the $s$-th encoder.
For each $j\in\J$,
let $\D_j$ be the index set of codewords available for the $j$-th decoder,
which is also the index set of messages reproduced by the decoder, where
$\D_j\subset\cS$.
Let $C_{\D_j,n}\equiv\{C_{s,n}\}_{s\in\D_j}$ be the set 
of codewords available for the $j$-th decoder,
$\Psi_{j,n}:\lrB{\Prod_{s\in\D_j}\C_{s,n}}\times\Y^n_j\to\Prod_{s\in\D_j}\Z_s^n$
be the $j$-th (possibly stochastic) decoding function,
and $\hZ_{\D_j}^n\equiv\Psi_{j,n}(C_{\D_j,n})$ be the reproduction of
messages by the $j$-th decoder.
For each $j\in\J$ and $s\in\D_j$,
let $\hZ^n_{j,s}$ be the reproduction of the $s$-th message by the $j$-th
decoder.
We expect $\hZ^n_{j,s}=Z^n_s$ for all $j\in\J$ and $s\in\D_j$
with a small error probability by letting $n$ be sufficiently large.
Let $\hZ^n_{\D_{\J}}\equiv\{\hZ^n_{j,s}\}_{j\in\J,s\in\D_j}$
be the random variable of all reproductions.

We call a rate vector
$\{r_s\}_{s\in\cS}$ {\em achievable} if
there is a (possibly stochastic) code
$\{(\{F_{s,n}\}_{s\in\cS}, \{\Psi_{j,n}\}_{j\in\J})\}_{n=1}^{\infty}$
such that
\begin{gather}
 \limsupn\frac{\log |\C_{s,n}|}n\leq r_s
 \quad\text{for all}\ s\in\cS
 \label{eq:ROPs-rate}
 \\
 \limn
 \Prob\lrsb{
  \hZ^n_{j,s}\neq Z^n_s\ \text{for some}\ j\in\J\ \text{and}\ s\in\D_j
 }
 =0,
 \label{eq:ROPs-error}
\end{gather}
where the joint distribution of
$(Z_{\cS}^n,C_{\cS,n},Y_{\J}^n,\hZ^n_{\D_{\J}})$ is given as
\begin{align*}
 &
 \mu_{Z_{\cS}^nC_{\cS,n}Y_{\J}^n\hZ^n_{\D_{\J}}}(\zz_{\cS},\cc_{\cS},\yy_{\J},\hzz_{\D_{\J}})
 \\*
 &\equiv
 \lrB{\prod_{j\in\J}\mu_{\hZ^n_{\D_j}|C_{\D_j,n}Y_j^n}(\hzz_{\D_j}|\cc_{\D_j},\yy_j)}
 \lrB{\prod_{s\in\cS}\mu_{C_{s,n}|Z_s^n}(\cc_s|\zz_s)}
 \notag
 \\*
 &\quad\cdot
 \mu_{Z_{\cS}^nY_{\J}^n}(\zz_{\cS},\yy_{\J})
\end{align*}
by using probability distributions
$\{\mu_{C_{s,n}|Z_s^n}\}_{s\in\cS}$
and $\{\mu_{\hZ^n_{\D_j}|C_{\D_j,n}Y_j^n}\}_{j\in\J}$,
corresponding to encoders and decoders, respectively.
Let $\ROPs$ be the set of all achievable rate vectors.

Let $\RITs$ be the set of all $\{r_s\}_{s\in\cS}$ satisfying
\begin{equation*}
 \sum_{s\in\D'_j}r_s
 \geq
 \oH(\ZZ_{\D'j}|\YY_j,\ZZ_{\D_j'^c})
\end{equation*}
for every $(j,\D_j')$ satisfying $j\in\J$ and
$\emptyset\neq\D_j'\subset\D_j$.
Then we have the following theorem,
which is a generalization of the results presented
in~\cite{C75}\cite{CSI82}\cite{MK95}\cite{HASH-BC}\cite{SWLDPC}\cite{SW73}.
\begin{thm}
$\ROPs=\RITs$.
\end{thm}

The converse part $\ROPs\subset\RITs$ is shown in
Appendix~\ref{sec:converse-source}.
To prove the achievability part $\ROPs\supset\RITs$,
we construct a code by assuming that $r_{\cS}\in\RITs$.
For each $s\in\cS$, a source $Z_s^n$ is encoded
by using a deterministic function $f_s:\Z_s^n\to\C_{s,n}$,
where we omit the dependence of $f_s$ on $n$
and $r_s=\log(|\C_{s,n}|)/n$ represents the encoding rate of the $s$-th
message.
We can use a sparse matrix as a function $f_s$ by assuming that
$\Z_s^n$ is an $n$-dimensional linear space on a finite field.

Let $\cc_s\in\C_{s,n}$ be the $s$-th codeword.
For each $j\in\J$, the decoder generates $\hZ_{\D_j}^n$
by using a constrained-random-number generator with a distribution
given as
\begin{align}
 &
 \mu_{\hZ_{\D_j}^n|C_{\D_j,n}Y_j^n}(\hzz_{\D_j}|\cc_{\D_j},\yy_j)
 \notag
 \\*
 &\equiv
 \frac{\mu_{Z_{\D_j}^n|Y_j^n}(\hzz_{\D_j}|\yy_j)\chi(f_{\D_j}(\hzz_{\D_j})=\cc_{\D_j})}
 {\mu_{Z_{\D_j}^n|Y_j^n}(\fC_{f_{\D_j}}(\cc_{\D_j})|\yy_j)}
 \label{eq:source-decoder}
\end{align}
for a given codeword $\cc_{\D_j}\equiv\{\cc_s\}_{s\in\D_j}$
and side information $\yy_j\in\Y_j^n$,
where
$f_{\D_j}(\hzz_{\D_j})\equiv\lrb{f_s(\hzz_s)}_{s\in\D_j}$.
It should be noted that
the constrained-random-number generator
is {\em sufficient} to achieve the fundamental limit.
When sources are i.i.d.,
tractable approximation algorithms for a
constrained-random-number
generator summarized in~\cite{SDECODING} are available.
While the maximum a posteriori probability decoder
is optimal, it may not be tractable.

Let $\Error(f_{\cS})$ be the decoding error probability
of a set of functions $f_{\cS}\equiv\{f_s\}_{s\in\cS}$.
Then we have the following theorem, which concludes the achievability part
$\ROPs\supset\RITs$. The proof is given in  Appendix~\ref{sec:proof-source}.
\begin{thm}
\label{thm:source}
Let $(\ZZ_{\cS},\YY_{\J})$ be a pair of general correlated sources.
Let us assume that $\{r_s\}_{s\in\cS}$ satisfies
\begin{align}
 \sum_{s\in\D'}r_s
 &>
 \oH(\ZZ_{\D_j'}|\YY_j,\ZZ_{\D_j'^c})
 \label{eq:rate-r}
\end{align}
for every $(j,\D_j')$ satisfying $j\in\J$ and $\emptyset\neq\D_j'\subset\D_j$.
Then there is a set of functions (sparse matrices)
$f_{\cS}$ such that
$\Error(f_{\cS})\leq\delta$
for any $\delta>0$ and all sufficiently large $n$.
\end{thm}

\section{Construction of Channel Code}

\subsection{General Formulas for Capacity Region}

Let $\cS$ be the index set of multiple messages,
$\I$ be the index set of channel inputs,
and $\J$ be the index set of channel outputs.
A general channel is characterized 
by a sequence $\{\mu_{Y^n_{\J}|X^n_{\I}}\}_{n=1}^{\infty}$
of conditional distributions,
where $X^n_{\I}\equiv\{X^n_i\}_{i\in\I}$
is a set of random variables of multiple channel inputs,
and $Y^n_{\J}\equiv\{Y^n_j\}_{j\in\J}$
is a set of random variables of multiple channel outputs.
For each $i\in\I$ and $n\in\NN$,
let $\X_i^n$ be the alphabet of random variable $X_i^n$.
For each $j\in\J$ and $n\in\NN$,
let $\Y_j^n$ be the alphabet of random variable $Y_j^n$.

For each $s\in\cS$ and $n\in\NN$,
let $M_{s,n}$ be a random variable of the $s$-th message
subject to the uniform distribution on an alphabet $\M_{s,n}$.
We assume that $\{M_{s,n}\}_{s\in\cS}$ are mutually independent.
For each $i\in\I$, 
let $\cS_i$ be the index set of sources available for the $i$-th encoder,
where $\cS_i\subset\cS$.
The $i$-th encoder generates the channel input $X^n_i$
from the set of messages $M_{\cS_i,n}\equiv\{M_{s,n}\}_{s\in\cS_i}$.
For each $j\in\J$,
let $\D_j$ be the index set of messages reproduced by the $j$-th
decoder, where $\D_j\subset\cS$.
The $j$-th decoder receives the channel output $Y^n_j$
and reproduces a set of messages
$\hM_{\D_j,n}\equiv\{\hM_{j,s,n}\}_{s\in\D_j}$,
where $\hM_{j,s,n}\in\M_{s,n}$
is the $s$-th message reproduced by the $j$-th encoder.
We expect that $\hM_{j,s,n}=M_{s,n}$ with a small error probability
for all $j\in\J$ and $s\in\D_j$ by letting $n$ be sufficiently large.

We call a rate vector
$\{R_s\}_{s\in\cS}$ {\em achievable} if
there is a (possibly stochastic) code
$\{(\{\Phi_{i,n}\}_{i\in\I},\{\Psi_{j,n}\}_{j\in\J})\}_{n=1}^{\infty}$
consisting of encoders
$\Phi_{i,n}:\Prod_{s\in\cS_i}\M_{s,n}\to\X^n_i$
and decoders $\Psi_{j,n}:\Y^n_j\to\Prod_{s\in\D_j}\M_{s,n}$
such that
\begin{gather}
 \liminfn\frac{\log |\M_{s,n}|}n\geq R_s
 \quad\text{for all}\ s\in\cS
 \label{eq:ROP-rate}
 \\
 \limn
 \Prob\lrsb{
  \hM_{j,s,n}\neq M_{s,n}\ \text{for some}\ j\in\J\ \text{and}\ s\in\D_j
 }
 =0,
 \label{eq:ROP-error}
\end{gather}
where $X_i^n\equiv\Phi_{i,n}(M_{\cS_i})$,
$\hM_{\D_j,n}\equiv\Psi_{j,n}(Y_j^n)$, and
the joint distribution of
$(M_{\cS,n},X_{\I}^n,Y_{\J}^n,\hM_{\D_{\J},n})$ is given as
\begin{align*}
 &
 \mu_{M_{\cS,n}X_{\I}^nY_{\J}^n\hM_{\D_{\J},n}}(\mm_{\cS},\xx_{\I},\yy_{\J},\hmm_{\D_{\J}})
 \notag
 \\*
 &
 =
 \lrB{\prod_{j\in\J}\mu_{\hM_{\D_j,n}|Y_{j}^n}(\hmm_{\D_j}|\yy_{j})}
 \mu_{Y_{\J}^n|X_{\I}^n}(\yy_{\J}|\xx_{\I})
 \notag
 \\*
 &\quad\cdot
 \lrB{\prod_{i\in\I}\mu_{X_i^n|M_{\cS_i,n}}(\xx_i|\mm_{\cS_i})}
 \lrB{\prod_{s\in\cS}\frac 1{|\M_{s,n}|}}
\end{align*}
by letting $\hM_{\D_{\J},n}\equiv\{\hM_{s,n}\}_{j\in\J,s\in\D_j}$.
Let $\ROP$ be the set of all achievable rate vectors.

Let $\RIT$  be defined as the set of all $\{R_s\}_{s\in\cS}$
satisfying the condition that
there are random variables $\{\ZZ_s\}_{s\in\cS}$
and positive numbers $\{r_s\}_{s\in\cS}$ such that
\begin{align}
 R_s
 &\geq 
 0
 \label{eq:RIT-R}
 \\
 \sum_{s\in\D_j'}r_s
 &\geq
 \oH(\ZZ_{\D'_j}|\YY_j,\ZZ_{\D_j'^c})
 \label{eq:RIT-r}
 \\
 R_s+r_s
 &\leq
 \uH(\ZZ_s)
 \label{eq:RIT-rR}
\end{align}
for all $(s,j,\D_j')$ satisfying $s\in\cS$, $j\in\J$, and
$\emptyset\neq\D_j'\subset\D_j$,
where the joint distribution of $(Z^n_{\cS},X^n_{\I},Y_{\J}^n)$ is given
as
\begin{align}
 &\mu_{Z_{\cS}^nX_{\I}^nY_{\J}^n}(\zz_{\cS},\xx_{\I},\yy_{\J})
 \notag
 \\*
 &=
 \mu_{Y_{\J}^n|X_{\I}^n}(\yy_{\J}|\xx_{\I})
 \lrB{\prod_{i\in\I}\mu_{X_i^n|Z^n_{\cS_i}}(\xx_i|\zz_{\cS_i})}
 \lrB{\prod_{s\in\cS}\mu_{Z^n_s}(\zz_s)}.
 \label{eq:joint-rit}
\end{align}
It should be noted that we can eliminate $\{r_s\}_{s\in\cS}$
from the above conditions
by employing the Fourier-Motzkin method~\cite[Appendix D]{EK11}.

We show the following theorem, which is a generalization of
the result presented in~\cite{CRNG}.
\begin{thm}
\label{thm:channel}
$\ROP = \RIT$.
\end{thm}
\begin{rem}
The capacity region of this type of channel is derived
in~\cite{SV06}
as the set of all
$\{R_s\}_{s\in\cS}$
that satisfy
\begin{align*}
 0\leq R_s
 \leq
 \min_{j:s\in\D_j}
 \uI(\ZZ_s;\YY_j)
 \quad\text{for all}\ s\in\cS,
\end{align*}
where $\uI(\UU;\VV)$ denotes the spectral inf-mutual information rate.
It should be noted that Theorem~\ref{thm:channel} provides an alternative
capacity region to that derived in~\cite{SV06}.
\end{rem}

The converse part $\ROP\subset\RIT$ will be shown in
Appendix~\ref{sec:converse}.
To prove the achievability part $\ROP\supset\RIT$,
we first consider a special case
where $\cS$ is the disjoint union of $\{\cS_i\}_{i\in\I}$.
We have the following theorem,
where the code construction is given in Section~\ref{sec:channel-code}.
\begin{thm}
\label{thm:disjoint}
Assume that $\cS$ is the disjoint union of $\{\cS_i\}_{i\in\I}$,
that is, $\cS=\bigcup_{i\in\I}\cS_i$ and $\cS_i\cap\cS_{i'}=\emptyset$
for all $i\neq i'$.
Let $\RITD$  be defined as the set of all
$\{R_s\}_{s\in\cS}$
satisfying the condition that
there are the random variables $\{\ZZ_s\}_{s\in\cS}$
and positive numbers $\{r_s\}_{s\in\cS}$
satisfying
(\ref{eq:RIT-R}), (\ref{eq:RIT-r}),
and
\begin{align}
 \sum_{s\in\cS_i'}[R_s+r_s]
 &\leq
 \uH(\ZZ_{\cS_i'})
 \label{eq:RIT-rR-disjoint}
\end{align}
for all $(s, i,\cS_i',j,\D_j')$
satisfying $s\in\cS$, $i\in\I$, $\emptyset\neq\cS_i'\subset\cS_i$,
$j\in\J$, and $\emptyset\neq\D_j'\subset\D_j$,
where the joint distribution of $(Z^n_{\cS},X^n_{\I},Y_{\J}^n)$ is given as
\begin{align}
 &
 \mu_{Z_{\cS}^nX_{\I}^nY_{\J}^n}(\zz_{\cS},\xx_{\I},\yy_{\J})
 \notag
 \\*
 &=
 \mu_{Y_{\J}^n|X_{\I}^n}(\yy_{\J}|\xx_{\I})
 \prod_{i\in\I}\mu_{X_i^nZ^n_{\cS_i}}(\xx_i,\zz_{\cS_i}).
 \label{eq:joint-rit-disjoint}
\end{align}
Then we have $\ROP\supset\RITD$.
\end{thm}

Next, by letting $\cS=\I$, $|\cS_i|=1$,
and $Z_i^n=X_i^n$ for each $i\in\I$,
we have the following corollary of Theorem~\ref{thm:disjoint}.
\begin{cor}
\label{thm:single}
Let $\{\mu_{Y^n_{\J}|Z^n_{\cS}}\}_{n=1}^{\infty}$ be a channel
with $|\cS|$ inputs.
Assume that the $s$-th encoder
has access to a single message $M_s^n$ for every $s\in\cS$.
Let $\RITS$  be defined as the set of all $\{R_s\}_{s\in\cS}$
satisfying the condition that
there are random variables $\{\ZZ_s\}_{s\in\cS}$
and positive numbers $\{r_s\}_{s\in\cS}$
satisfying (\ref{eq:RIT-R})--(\ref{eq:RIT-rR}) for all $(s,j,\D_j')$
satisfying $s\in\cS$, $j\in\J$, and $\emptyset\neq\D_j'\subset\D_j$,
where the joint distribution of $(Z^n_{\cS},Y_{\J}^n)$ is given as
\begin{align*}
 \mu_{Z_{\cS}^nY_{\J}^n}(\zz_{\cS},\yy_{\J})
 &=
 \mu_{Y_{\J}^n|Z_{\cS}^n}(\yy_{\J}|\zz_{\cS})
 \prod_{s\in\cS}\mu_{Z_s^n}(\zz_s).
\end{align*}
Then we have $\ROP\supset\RITS$.
\end{cor}

By using this corollary, we show the achievability part $\ROP\supset\RIT$ of
Theorem~\ref{thm:channel}.
In the following, we assume that
for given $\mu_{Y_{\J}^n|X_{\I}^n}$, 
$\{\mu_{X^n_i|Z_{\cS_i}^n}\}_{i\in\I}$,
and $\{\mu_{Z_s^n}\}_{s\in\cS}$, 
the rate vector $\{R_s\}_{s\in\cS}$ satisfies
$\{R_s\}_{s\in\cS}\in\RIT$.
For a given channel
$\mu_{Y^n_{\J}|X^n_{\I}}$, let us consider an $|\cS|$-input channel
$\mu_{Y^n_{\J}|Z^n_{\cS}}$ defined as
\begin{align}
 \mu_{Y_{\J}^n|Z_{\cS}^n}(\yy_{\J}|\zz_{\cS})
 &\equiv
 \sum_{\xx_{\I}}
 \mu_{Y_{\J}^n|X_{\I}^n}(\yy_{\J}|\xx_{\I})
 \prod_{i\in\I}\mu_{X_i^n|Z^n_{\cS_i}}(\xx_i|\zz_{\cS_i}).
 \label{eq:muYgU}
\end{align}
We reduce the scenario of multiple common messages for the channel
$\mu_{Y_{\cS}^n|\X_{\I}^n}$
to the scenario for a channel $\mu_{Y_{\cS}^n|Z_{\cS}^n}$
in which the $s$-th input terminal has access to a single message
$M_{s,n}$. 

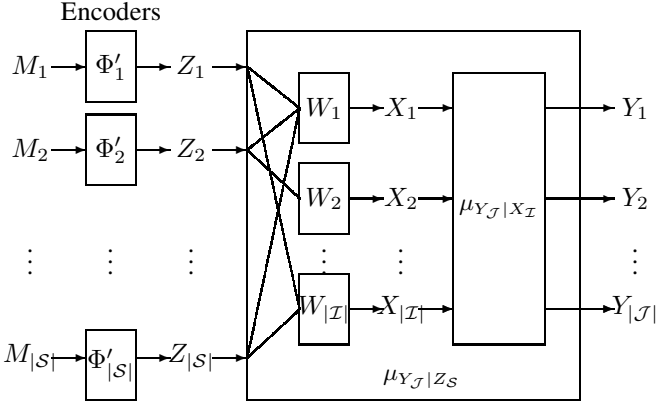
\begin{figure}
\begin{center}
 \unitlength 0.46mm
 \begin{picture}(180,120)(-62,-6)
  \put(-34,106){\makebox(0,0){Encoders}}
  \put(-57,90){\makebox(0,0){$M_1$}}
  \put(-57,66){\makebox(0,0){$M_2$}}
  \put(-57,36){\makebox(0,0){$\vdots$}}
  \put(-57,6){\makebox(0,0){$M_{|\cS|}$}}
  \put(-51,90){\vector(1,0){10}}
  \put(-51,66){\vector(1,0){10}}
  \put(-51,6){\vector(1,0){10}}
  \put(-41,80){\framebox(14,20){$\Phi'_1$}}
  \put(-41,56){\framebox(14,20){$\Phi'_2$}}
  \put(-34,36){\makebox(0,0){$\vdots$}}
  \put(-41,-6){\framebox(14,20){$\Phi'_{|\cS|}$}}
  \put(-27,90){\vector(1,0){10}}
  \put(-27,66){\vector(1,0){10}}
  \put(-27,6){\vector(1,0){10}}
  \put(-11,90){\makebox(0,0){$Z_1$}}
  \put(-11,66){\makebox(0,0){$Z_2$}}
  \put(-11,36){\makebox(0,0){$\vdots$}}
  \put(-11,6){\makebox(0,0){$Z_{|\cS|}$}}
  \put(-5,90){\vector(1,0){10}}
  \put(-5,66){\vector(1,0){10}}
  \put(-5,6){\vector(1,0){10}}
  \qbezier(5,90)(12.5,84)(20,78)
  \qbezier(5,90)(12.5,55)(20,20)
  \qbezier(5,66)(12.5,72)(20,78)
  \qbezier(5,66)(12.5,59)(20,52)
  \qbezier(5,6)(12.5,13)(20,20)
  \qbezier(5,6)(12.5,42)(20,78)
  \put(20,68){\framebox(14,20){$W_1$}}
  \put(20,42){\framebox(14,20){$W_2$}}
  \put(27,36){\makebox(0,0){$\vdots$}}
  \put(20,10){\framebox(14,20){$W_{|\I|}$}}
  
  \put(54,20){\vector(1,0){10}}
  \put(34,78){\vector(1,0){10}}
  \put(49,78){\makebox(0,0){$X_1$}}
  \put(54,78){\vector(1,0){10}}
  \put(34,52){\vector(1,0){10}}
  \put(49,52){\makebox(0,0){$X_2$}}
  \put(54,52){\vector(1,0){10}}
  \put(49,36){\makebox(0,0){$\vdots$}}
  \put(34,20){\vector(1,0){10}}
  \put(49,20){\makebox(0,0){$X_{|\I|}$}}
  \put(54,20){\vector(1,0){10}}
  \put(64,10){\framebox(26,78){\small $\mu_{Y_{\J}|X_{\I}}$}}
  \put(5,-6){\framebox(95,106){}}
  \put(55,0){\makebox(0,0){\small $\mu_{Y_{\J}|Z_{\cS}}$}}
  
  \put(90,20){\vector(1,0){20}}
  \put(116,78){\makebox(0,0){$Y_1$}}
  \put(90,78){\vector(1,0){20}}
  \put(116,52){\makebox(0,0){$Y_2$}}
  \put(90,52){\vector(1,0){20}}
  \put(116,36){\makebox(0,0){$\vdots$}}
  \put(116,20){\makebox(0,0){$Y_{|\J|}$}}
  \put(90,20){\vector(1,0){20}}
 \end{picture}
\end{center}
\caption{Reduction of Multiple Common Messages to Private Messages:
 Decoders are the same as Fig.~\ref{fig:channel}}
\label{fig:reduction}
\end{figure}

\begin{figure}
\begin{center}
 \unitlength 0.62mm
 \begin{picture}(126,100)(0,4)
  \put(65,97){\makebox(0,0){Encoder $\Phi_i$}}
  \put(5,76){\makebox(0,0){$\mm_{s_1}$}}
  \put(5,52){\makebox(0,0){$\mm_{s_2}$}}
  \put(5,38){\makebox(0,0){$\vdots$}}
  \put(5,20){\makebox(0,0){$\mm_{s_{|\cS_i|}}$}}
  \put(14,76){\vector(1,0){20}}
  \put(14,52){\vector(1,0){20}}
  \put(14,20){\vector(1,0){20}}
  \put(34,66){\framebox(14,20){$\Phi'_{s_1}$}}
  \put(34,42){\framebox(14,20){$\Phi'_{s_2}$}}
  \put(41,38){\makebox(0,0){$\vdots$}}
  \put(34,10){\framebox(14,20){$\Phi'_{s_{|\cS_i|}}$}}
  \put(48,76){\vector(1,0){10}}
  \put(48,52){\vector(1,0){10}}
  \put(48,20){\vector(1,0){10}}
  \put(65,76){\makebox(0,0){$\zz_{s_1}$}}
  \put(65,52){\makebox(0,0){$\zz_{s_2}$}}
  \put(65,38){\makebox(0,0){$\vdots$}}
  \put(65,20){\makebox(0,0){$\zz_{s_{|\cS_i|}}$}}
  \put(72,76){\vector(1,0){10}}
  \put(72,52){\vector(1,0){10}}
  \put(72,20){\vector(1,0){10}}
  \put(82,10){\framebox(14,76){$W_i$}}
  \put(96,48){\vector(1,0){20}}
  \put(121,48){\makebox(0,0){$\xx_i$}}
  \put(24,4){\framebox(82,88){}}
 \end{picture}
\end{center}
\caption{Construction of $i$-th Encoder}
\label{fig:code-i}
\end{figure}
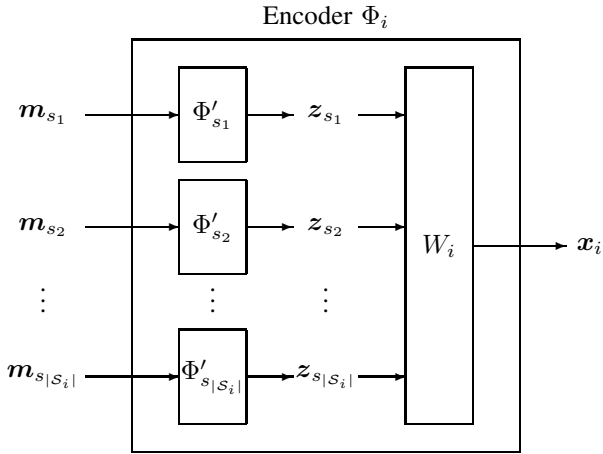

Since conditions (\ref{eq:RIT-R})--(\ref{eq:RIT-rR})
depend only on the joint distribution of $(\ZZ_{\cS},\YY_{\J})$,
we can apply Corollary~\ref{thm:single}
to the channel $\mu_{Y_{\J}^n|Z_{\cS}^n}$ defined by (\ref{eq:muYgU}).
We have the fact that
there is a code $(\Phi'_{\cS,n},\Psi'_{\J,n})$ for this channel
at $\{R_s\}_{s\in\cS}\in\RIT$.
Figure \ref{fig:reduction} illustrates the code construction
for the channel $\mu_{Y_{\J}^n|Z_{\cS}^n}$.
The code  $(\Phi_{\I,n},\Psi_{\J,n})$
for the channel $\mu_{Y_{\J}^n|X^n_{\I}}$ is given as
\begin{align*}
 \Phi_{i,n}(\mm_{\cS_i})
 &\equiv
 W_i\lrsb{\{\Phi'_s(\mm_s)\}_{s\in\cS_i}}
 \\
 \Psi_{j,n}(\yy_j)
 &\equiv
 \Psi'_{j,n}(\yy_j)
\end{align*}
for a multiple message $\mm_{\cS}$,
where $W_i$ is a stochastic function subject to the distribution
$\mu_{X^n_i|Z_{\cS_i}^n}$ for each $i\in\I$.
Figure \ref{fig:code-i} illustrates the construction of the $i$-th
encoder.
Since the above discussion implies the achievability part
$\ROP\supset\RIT$,
the proof of Theorem~\ref{thm:channel} is completed.

\subsection{Code Construction}
\label{sec:channel-code}

This section introduces a channel code for the proof of
Theorem~\ref{thm:disjoint}.
The idea for the construction is drawn from \cite{CRNG}\cite{HASH}\cite{SW2CC}.

For a pair of correlated sources $(\ZZ_{\cS},\YY_{\J})$,
we use the multi-terminal source code introduced in Section~\ref{sec:source},
where $f_s$ is the $s$-th deterministic encoder and
$\hZ^n_{\D_j}$ is the output of the $j$-th stochastic decoder.
Let $r_s=\log(|\im f_s|)/n$ be the encoding rate of the $s$-th encoder
and $\Error(f_{\cS})$ be the decoding error of this code.
It should be noted that
it is unnecessary for $\Error(f_{\cS})$ to be close to zero.

For each $s\in\cS$ and $n$,
let $g_s:\Z^n_s\to\M_{s,n}$ be a function,
where $R_s\equiv\log(|\M_{s,n}|)/n$ represents the rate of the $s$-th message.
We can use a sparse matrix as a function $g_s$ by assuming that
$\Z_s^n$ is an $n$-dimensional linear space on a finite field.
Let $\cc_s\in\C_{s,n}$ be a vector that is generated at random
subject to the distribution
$\{\mu_{Z_s^n}(\fC_{f_s}(\cc_s))\}_{\cc_s\in\im f_s}$.
We can obtain $\cc_s$ by generating
$\zz_s\in\Z_s^n$ at random subject to the distribution $\mu_{Z_s^n}$
and letting $\cc_s\equiv f_s(\zz_s)$.

In addition to $f_{\cS}\equiv\{f_s\}_{s\in\cS}$,
we fix a set of functions
$g_{\cS}\equiv\{g_s\}_{s\in\cS}$ and a set of vectors
$\cc_{\cS}\equiv\{\cc_s\}_{s\in\cS}$
so that they are available for constructing an encoder and a decoder.
For each $i\in\I$,
the $i$-th encoder uses $f_{\cS_i}$, $g_{\cS_i}$, and $\cc_{\cS_i}$.
For each $j\in\J$,
the $j$-th decoder uses $\cc_{\D_j}$ and $g_{\D_j}$
as well as the decoder of source code,
where $f_{\D_j}$ is used implicitly.
We fix the probability distributions
$\{\mu_{Z_{\cS_i}^n}\}_{i\in\I}$ and conditional probability distributions
$\{\mu_{X_i^n|Z^n_{\cS_i}}\}_{i\in\I}$.

Here, we define a constrained-random-number generator,
which is used by the $i$-th encoder.
Let $\tZ_i^n$
be a random variable corresponding to the distribution
\begin{align*}
 &
 \mu_{\tZ_i^n|C_{\cS_i,n}M_{\cS_i,n}}(\zz_{\cS_i}|\cc_{\cS_i},\mm_{\cS_i})
 \notag
 \\*
 &\equiv
 \frac{
  \mu_{Z_{\cS_i}^n}(\zz_{\cS_i})
  \chi(f_{\cS_i}(\zz_{\cS_i})=\cc_{\cS_i},g_{\cS_i}(\zz_{\cS_i})=\mm_{\cS_i})
 }{
  \mu_{Z_{\cS_i}^n}(\fC_{f_{\cS_i}}(\cc_{\cS_i})\cap\fC_{g_{\cS_i}}(\mm_{\cS_i}))
 }.
\end{align*}
We define the $i$-th encoder
$\Encoder_{i,n}:\M_{\cS_i,n}\to\X_i^n$
and the $j$-th decoder $\Decoder_{j,n}:\Y_j^n\to\M_{\D_j,n}$
as
\begin{align*}
 \Phi_{i,n}(\mm_{\cS_i})
 &\equiv
 W_i(\tZ^n_{\cS_i})
 \\
 \Psi_{j,n}(\yy_j)
 &\equiv
 \{g_s(\hZ^n_{j,s})\}_{s\in\D_j},
\end{align*}
where the encoder claims an error when
$\fC_{f_{\cS_i}}(\cc_{\cS_i})\cap\fC_{g_{\cS_i}}(\mm_{\cS_i})=\emptyset$,
$W_i$ is the channel subject to
the conditional probability distribution
$\mu_{X_i|Z^n_{\cS_i}}$, and $\hZ^n_{j,s}$
is the projection of $\hZ^n_{\D_j}$ on $\M_{s,n}$.
The flow of vectors is illustrated in
Fig.\ \ref{fig:channel-code} in Appendix~\ref{sec:proof-channel}.

Let $\hM_{\D_j,n}\equiv \Psi_j(Y^n_j)$
and $\Error(f_{\cS},g_{\cS},\cc_{\cS})$ 
be the error probability.
We have the following lemma, where the proof is given in
Appendix~\ref{sec:proof-channel}.
\begin{thm}
\label{lem:channel-encoder}
Let us assume that $\cS$ is the disjoint union of $\{\cS_i\}_{i\in\I}$,
that is, $\cS=\bigcup_{i\in\I}\cS_i$ and $\cS_i\cap\cS_{i'}=\emptyset$
for all $i\neq i'$.
Let us assume that $\{(r_s,R_s)\}_{s\in\cS}$ satisfies
\begin{align}
 \sum_{s\in\cS'_i}[R_s+r_s]
 &<
 \uH(\ZZ_{\cS'_i})
 \label{eq:channel-encoder}
\end{align}
for all $(i,\cS_i')$ satisfying $i\in\I$ and $\emptyset\neq\cS_i'\subset\cS_i$.
Then for any $\delta>0$
and all sufficiently large $n$
there are functions (sparse matrices)
$g_{\cS}$ and vectors $\cc_{\cS}$ such that
$\Error(f_{\cS},g_{\cS},\cc_{\cS})\leq\Error(f_{\cS})+\delta$.
\end{thm}

Immediately from Theorems~\ref{thm:source} and~\ref{lem:channel-encoder},
we have Theorem~\ref{thm:disjoint}.

\section{Application to Multiple Access Channel}

In this section, we apply code construction
to a multiple access channel,
where $\cS\equiv\{0,1,2\}$, $\I\equiv\{1,2\}$,
$\cS_i\equiv\{0,i\}$ for each $i\in\I$,
$\J\equiv\{0\}$, and $\D_0\equiv\{1,2\}$.
This setting corresponds to the situation
that the first and the second encoder have access to the $0$-th message.
In the following, we denote $\D\equiv\D_0$, $\YY\equiv\YY_0$,
$Y^n\equiv Y^n_0$, and $\yy\equiv\yy_0$.
From Theorem~\ref{thm:channel},
we have an achievable region as the set of
all $\{R_s\}_{s\in\cS}$ such that there are $\ZZ_{\cS}$ and
positive numbers $\{r_s\}_{s\in\cS}$ satisfying
{\small\begin{align*}
  R_s
  &\geq 0
  \\
  \sum_{s\in\D'} r_s
  &\geq
  \oH(\ZZ_{\D'}|\YY,\ZZ_{\D'^c})
  \\
  R_s+r_s
  &\leq
  \uH(\ZZ_s)
\end{align*}}%
for all $(s,\D')$ satisfying $s\in\cS$ and $\emptyset\neq\D'\subset\D$.
where the joint distribution of
$(Z^n_{\cS},X^n_{\I},Y^n)$ is given
by (\ref{eq:joint-rit}).

Here, let us assume that a multiple access channel is
i.i.d.
given by a conditional distribution $\mu_{Y|X_1X_2}$.
In addition, let us assume that
$(\ZZ_0,\ZZ_1,\ZZ_2,\XX_1,\XX_2)$ are i.i.d.\ sources
given by the distributions $\mu_{Z_0}$, $\mu_{Z_1}$, $\mu_{Z_2}$,
$\mu_{X_1|Z_0Z_1}$, and $\mu_{X_2|Z_0Z_2}$.
By employing the Fourier-Motzkin method~\cite[Appendix D]{EK11}
to eliminate $\{r_s\}_{s\in\cS}$,
we have following equivalent conditions for $(R_0,R_1,R_2)$ as
{\small\begin{align*}
  0\leq R_0
  &\leq
  I(Z_0;Y|Z_1,Z_2)
  \\
  0\leq R_1
  &\leq
  I(Z_1;Y|Z_0,Z_2)
  \\
  0\leq R_2
  &\leq
  I(Z_2;Y|Z_0,Z_1)
  \\
  R_0+R_1
  &\leq
  I(Z_0,Z_1;Y|Z_2)
  \\
  R_0+R_2
  &\leq
  I(Z_0,Z_2;Y|Z_1)
  \\
  R_1+R_2
  &\leq
  I(Z_1,Z_2;Y|Z_0)
  \\
  R_0+R_1+R_2
  &\leq
  I(Z_0,Z_1,Z_2;Y),
\end{align*}}%
where we use
the fact that $Z_0$, $Z_1$, and $Z_2$ are mutually independent.
By making the convex closure after the union over all
i.i.d.\ distributions $\mu_{Z_0}$, $\mu_{Z_1}$, $\mu_{Z_2}$,
$\mu_{X_1|Z_0Z_1}$, and $\mu_{X_2|Z_0Z_2}$,
we have the region equivalent to that derived in \cite{H79}.
It is shown in \cite{H79} that
this region is equivalent to the region derived in \cite{SW73MAC}
specified by the conditions
{\small\begin{align*}
  0\leq R_1
  &\leq
  I(X_1;Y|X_2,U)
  \\
  0\leq R_2
  &\leq
  I(X_2;Y|X_2,U)
  \\
  R_1+R_2
  &\leq
  I(X_1,X_2;Y|U),
  \\
  R_0+R_1+R_2
  &\leq
  I(X_1,X_2;Y),
\end{align*}}%
where the joint distribution of $(U,X_1,X_2,Y)$ is given as
\begin{align*}
 &\mu_{UX_1X_2Y}(u,x_1,x_2,y)
 \\*
 &\equiv
 \mu_{Y|X_1X_2}(y|x_1,x_2)
 \mu_{X_1|U}(x_1|u)
 \mu_{X_2|U}(x_2|u)
 \mu_{U}(u).
\end{align*}

\section{Application to Broadcast Channel}

In this section, we apply code construction
to a broadcast channel,
where $\I\equiv\{0\}$, $\cS\equiv\cS_0\equiv\{0,1,2\}$, $\J\equiv\{1,2\}$,
and $\D_j\equiv\{0,j\}$ for each $j\in\J$.
This setting corresponds to the situation
that both decoders reproduce the $0$-th message.
In the following, we denote 
$\XX\equiv\XX_0$, $X^n\equiv X^n_0$, and $\xx\equiv\xx_0$.
From Theorem~\ref{thm:disjoint},
we have an achievable region of
all $\{R_s\}_{s\in\cS}$ such that there are
$\ZZ_{\cS}$ and $\{r_s\}_{s\in\cS}$ satisfying
{\small\begin{align*}
  R_s
  &\geq 0
  \\
  \sum_{s\in\D_j'} r_s
  &\geq
  \oH(\ZZ_{\D_j'}|\YY_j,\ZZ_{\D_j'^c})
  \\
  \sum_{s\in\cS'} \lrB{R_s+r_s}
  &\leq
  \uH(\ZZ_{\cS'})
\end{align*}}%
for all $(s,\cS',j,\D_j')$
satisfying $s\in\cS$, $\emptyset\neq\cS'\subset\cS$,
$j\in\J$, and $\emptyset\neq\D_j'\subset\D_j$.
where the joint distribution of
$(Z^n_{\cS},X^n,Y_{\J}^n)$ is given
by (\ref{eq:joint-rit-disjoint}).

Here, let us assume that a broadcast channel is
i.i.d.\ given by a conditional distribution $\mu_{Y_0Y_1|X}$.
In addition, let us assume that
$(\XX,\ZZ_0,\ZZ_1,\ZZ_2)$ are i.i.d.\ sources
given by the distribution $\mu_{XZ_0Z_1Z_3}$.
By employing the Fourier-Motzkin method~\cite[Appendix D]{EK11}
to eliminate $\{r_s\}_{s\in\cS}$
and remove redundant conditions, we have
an inner region of $\ROP$ specified by the following conditions
for $(R_0,R_1,R_2)$ as
{\small
 \begin{align}
  &0\leq R_0
  \leq
  \min\{I(Z_0;Z_1Y_1),I(Z_0;Z_2Y_2)\}
  \notag
  \\
  &0\leq R_1
  \leq I(Z_1;Z_0Y_1)
  \notag
  \\
  &0\leq R_2
  \leq I(Z_2;Z_0Y_2)
  \notag
  \\
  &R_0+R_1
  \leq
  I(Z_1;Y_1|Z_0)+\min\{I(Z_0;Y_1),I(Z_0;Z_2Y_2)\}
  \notag
  \\
  &R_0+R_2
  \leq
  I(Z_2;Y_2|Z_0)+\min\{I(Z_0;Y_2),I(Z_0;Z_1Y_1)\}
  \notag
  \\
  &R_1+R_2
  \leq
  I(Z_1;Z_0Y_1)+I(Z_2;Z_0Y_2)-I(Z_1;Z_2)
  \notag
  \\
  &R_0+R_1+R_2
  \leq
  I(Z_1;Y_1|Z_0)+I(Z_2;Y_2|Z_0)
  -I(Z_1;Z_2|Z_0)
  \notag
  \\*
  &\qquad\qquad\qquad\qquad
  +\min\{I(Z_0;Y_1),I(Z_0;Y_2)\}
  \notag
  \\
  &2R_0+R_1+R_2
  \leq
  I(Z_0Z_1;Y_1)+I(Z_0Z_2;Y_2)
  -I(Z_1;Z_2|Z_0).
  \label{eq:bc-R2012}
\end{align}}%
By making the convex closure after the union over all i.i.d.\ distributions
$\mu_{XZ_0Z_1Z_3}$,
we have the fact that this region
is equivalent to the Marton inner
region~\cite[Problem 16.10 (c)]{CK11} specified by the conditions
{\small
 \begin{align}
  R_i&\geq0\quad\text{for}\ i\in\{0,1,2\}
  \label{eq:bc-Rpositive}
  \\
  R_0&\leq\min\{I(Z_0;Y_1),I(Z_0;Y_2)\}
  \label{eq:bc-martonR0}
  \\
  R_0+R_1&\leq I(Z_0Z_1;Y_1)
  \label{eq:bc-martonR01}
  \\
  R_0+R_2&\leq I(Z_0Z_2;Y_2)
  \label{eq:bc-martonR02}
  \\
  R_0+R_1+R_2
  &
  \leq I(Z_1;Y_1|Z_0)+I(Z_2;Y_2|Z_0)-I(Z_1;Z_2|Z_1)
  \notag
  \\*
  &\quad
  +\min\{I(Z_0;Y_1),I(Z_0;Y_2)\},
  \label{eq:bc-martonR012}
\end{align}}%
where the equivalence comes from the result presented in~\cite{ITW13}.
It should noted that this region is also equivalent
to the Gel'fand-Pinsker region~\cite{GP80}
specified by inequalities
(\ref{eq:bc-Rpositive}),
(\ref{eq:bc-martonR0}),
(\ref{eq:bc-martonR012}),
and
{\small
 \begin{align*}
  R_0+R_1
  &\leq
  I(Z_1;Y_1|Z_0)+\min\{I(Z_0;Y_1),I(Z_0;Y_2)\}
  \\
  R_0+R_2
  &\leq
  I(Z_2;Y_2|Z_0)+\min\{I(Z_0;Y_2),I(Z_0;Y_1)\},
\end{align*}}%
and the Liang-Kramer-Poor region~\cite{LKP11} specified by inequalities
(\ref{eq:bc-R2012}), (\ref{eq:bc-Rpositive}), and (\ref{eq:bc-martonR01})--(\ref{eq:bc-martonR012}).

\appendix

\subsection{Entropy and Mutual Information for General Sources}
\label{sec:ispec}

First, we review the definition of the limit superior/inferior in
probability introduced in \cite{HAN}.
For a sequence $\{U_n\}_{n=1}^{\infty}$ of random variables,
the {\em limit superior in probability} $\plimsupn U_n$
and the {\em limit inferior in probability} $\pliminfn U_n$ are defined
as
\begin{align*}
 \plimsupn U_n
 &\equiv 
 \inf\lrb{\theta: \limn \Prob\lrsb{U_n>\theta}=0}
 \\
 \pliminfn U_n
 &\equiv
 \sup\lrb{\theta: \limn \Prob\lrsb{U_n<\theta}=0}.
\end{align*}
We have the following relations~\cite[Section 1.3]{HAN}:
\begin{align}
 \plimsupn\lrB{U_n+V_n}
 &\leq \plimsupn U_n + \plimsupn V_n
 \label{eq:plimsup-upper}
 \\
 \plimsupn\lrB{U_n+V_n}
 &\geq \plimsupn U_n + \pliminfn V_n
 \label{eq:plimsup-lower}
 \\
 \pliminfn\lrB{U_n+V_n}
 &\leq \plimsupn U_n + \pliminfn V_n
 \label{eq:pliminf-upper}
 \\
 \pliminfn\lrB{U_n+V_n}
 &\geq \pliminfn U_n + \pliminfn V_n
 \label{eq:pliminf-lower}
 \\
 \plimsupn\lrB{-U_n}
 &=\pliminfn U_n.
 \label{eq:plimsup-pliminf}
\end{align}

For a sequence $\{\mu_{U^n}\}_{n=1}^{\infty}$
of probability distributions corresponding to $\UU$,
we define the spectral inf-entropy rate $\uH(\UU)$ as
\begin{align*}
 \uH(\UU)
 &\equiv
 \pliminfn\frac 1n\log_2\frac1{\mu_{U^n}(U^n)}.
\end{align*}

For a general sequence $\{\mu_{U^nV^n}\}_{n=1}^{\infty}$ of
joint probability distributions
corresponding to $(\UU,\VV)=\{(U^n,V^n)\}_{n=1}^{\infty}$,
we define the spectral conditional sup-entropy rate $\oH(\UU|\VV)$,
the spectral conditional inf-entropy rate $\uH(\UU|\VV)$,
and the spectral inf-information rate
$\uI(\UU;\VV)$
as
\begin{align*}
 \oH(\UU|\VV)
 &\equiv
 \plimsupn\frac 1n\log_2\frac1{\mu_{U^n|V_n}(U^n|V^n)}
 \\
 \uH(\UU|\VV)
 &\equiv
 \pliminfn\frac 1n\log_2\frac1{\mu_{U^n|V_n}(U^n|V^n)}
 \\
 \uI(\UU;\VV)
 &\equiv
 \pliminfn\frac 1n\log_2\frac{\mu_{U^n|V^n}(U^n|V^n)}{\mu_{U^n}(U^n)}.
\end{align*}

In the following, we introduce some inequalities that we use in the
proof of the converse part.
Trivially, we have
\begin{equation*}
 \oH(\UU|\VV)\geq\uH(\UU|\VV)\geq 0.
\end{equation*}
From \cite[Lemma 3.2.1, Definition 4.1.3]{HAN},
we have
\begin{equation}
 \pliminfn\frac 1n\log_2\frac{\mu_{U_n}(U_n)}{\mu_{V_n}(U_n)}\geq 0,
 \label{eq:pliminf-div}
\end{equation}
which implies that
\begin{align}
 \uI(\UU;\VV)
 &
 =
 \pliminfn\frac 1n\log_2
 \frac{\mu_{U^n|V^n}(U^n|V^n)}
 {\mu_{U^n}(U^n)}
 \notag
 \\
 &=
 \pliminfn\frac 1n\log_2
 \frac{\mu_{U^nV^n}(U^n,V^n)}
 {\mu_{U^n}(U^n)\mu_{V^n}(V^n)}
 \notag
 \\
 &
 \geq 0.
\end{align}
We show the following lemmas.
\begin{lem}
Let $\U_n$ be the alphabet of $U^n$. Then
\begin{equation*}
 \oH(\UU)
 \leq
 \limsupn\frac{\log_2|\U_n|}n.
\end{equation*}
\end{lem}
\begin{IEEEproof}
 We have
 \begin{align}
  &
  \limsupn\frac{\log_2|\U_n|}n-\oH(\UU)
  \notag
  \\*
  &=
  \plimsupn\frac{\log_2|\U_n|}n-\plimsupn\frac1n\log_2\frac1{\mu_{U^n}(U^n)}
  \notag
  \\
  &=
  \plimsupn\frac{\log_2|\U_n|}n+\pliminfn\frac1n\log_2\mu_{U^n}(U^n)
  \notag
  \\
  &\geq
  \pliminfn\frac 1n\log_2(|\U_n|\mu_{U^n}(U^n))
  \notag
  \\
  &=
  \pliminfn\frac 1n\log_2\frac{\mu_{U^n}(U^n)}{1/|\U_n|}
  \notag
  \\
  &\geq
  0,
 \end{align}
 where
 the second equality comes from (\ref{eq:plimsup-pliminf}),
 the first inequality comes from (\ref{eq:pliminf-upper}),
 and the second inequality comes from (\ref{eq:pliminf-div}) by letting
 $\mu_{V_n}$ be the uniform distribution on $\U_n$.
\end{IEEEproof}
\begin{lem}
$\oH(\UU|\VV)\geq \oH(\UU|\VV,\WW)$.
\end{lem}
\begin{IEEEproof}
 We have
 \begin{align}
  &
  \oH(\UU|\VV)-\oH(\UU|\VV,\WW)
  \notag
  \\*
  &=
  \plimsupn\frac 1n\log_2
  \frac1{\mu_{U^n|V^n}(U^n|V^n)}
  \notag
  \\*
  &\quad
  -
  \plimsupn\frac 1n\log_2\frac1{\mu_{U^n|V^nW^n}(U^n|V^n,W^n)}
  \notag
  \\*
  &=
  \plimsupn\frac 1n\log_2
  \frac1{\mu_{U^n|V^n}(U^n|V^n)}
  \notag
  \\*
  &\quad
  +
  \pliminfn\frac 1n\log_2\mu_{U^n|V^nW^n}(U^n|V^n,W^n)
  \notag
  \\
  &\geq
  \pliminfn\frac 1n\log_2
  \frac{\mu_{U^n|V^nW^n}(U^n|V^n,W^n)}{\mu_{U^n|V^n}(U^n|V^n)}
  \notag
  \\
  &=
  \pliminfn\frac 1n\log_2
  \frac{\mu_{U^nV^nW^n}(U^n,V^n,W^n)}
  {\mu_{U^n|V^n}(U^n|V^n)\mu_{V^nW^n}(V^n,W^n)}
  \notag
  \\
  &\geq
  0,
 \end{align}
 where the second equality comes from (\ref{eq:plimsup-pliminf}),
 the first inequality comes from (\ref{eq:pliminf-upper}),
 and the second inequality comes from (\ref{eq:pliminf-div}).
\end{IEEEproof}

\subsection{Proof of $\ROPs\subset\RITs$}
\label{sec:converse-source}
We use the following lemma, which is analogous
to the Fano inequality.
\begin{lem}[{\cite[Lemma 7]{CRNG}}]
\label{lem:fano}
Let $(\UU,\VV)\equiv\{(U_n,V_n)\}_{n=1}^{\infty}$ be a sequence
of two random variables.
If there is a sequence $\{\Psi_n\}_{n=1}^{\infty}$
of (possibly stochastic) functions independent of $(\UU,\VV)$
satisfying the condition
\begin{equation*}
 \limn \Prob(\Psi_n(V_n)\neq U_n)=0,
\end{equation*}
then
\begin{equation*}
 \oH(\UU|\VV)=0.
\end{equation*}
\end{lem}
\begin{IEEEproof}
When $\{\Psi_n\}_{n=1}^{\infty}$ is a sequence of deterministic
functions, the lemma is the same as \cite[Lemma 7]{CRNG}.
When $\{\Psi_n\}_{n=1}^{\infty}$ is a sequence of stochastic functions,
we can obtain a sequence $\{\psi_n\}_{n=1}^{\infty}$ of deterministic functions
such that
\begin{align*}
 \Prob(\psi_n(V_n)\neq U_n)
 &\leq
 \sum_{\psi_n}\Prob(\Psi_n=\psi_n)\Prob(\psi_n(V_n)\neq U_n)
 \\
 &=
 \Prob(\Psi_n(V_n)\neq U_n)
\end{align*}
for all $n$ from the random coding argument
and the fact that $\Psi_n$ is independent of $(U_n,V_n)$.
Then we have the lemma by using~\cite[Lemma 7]{CRNG}.
\end{IEEEproof}

In the following, we show $\ROPs\subset\RITs$ by using the above lemma.

Assume that $\{r_s\}_{s\in\cS}\in\ROPs$.
Then there is a code
$\{(\{F_{s,n}\}_{s\in\cS},\{\Psi_{j,n}\}_{j\in\J})\}_{n=1}^{\infty}$
satisfying (\ref{eq:ROPs-rate}) and (\ref{eq:ROPs-error}).

For $j\in\J$ and $\D_j'\subset\D_j$,
let
$\Psi_{j,\D_j',n}(C_{\D_j},Y^n_j)$
be the projection of $\Psi_{j,n}(C_{\D_j},Y^n_j)$ on
$\Prod_{s\in\D_j'}\Z_s^n$.
Then we have
\begin{equation*}
 \limn\Prob(\Psi_{j,\D_j',n}(C_{\D_j,n},Y_j^n)\neq Z_{\D_j'}^n)=0
\end{equation*}
from (\ref{eq:ROPs-error}).
From Lemma~\ref{lem:fano},
we have
\begin{align}
 \oH(\ZZ_{\D_j'}|\CC_{\D_j'},\CC_{\D_j'^c},\YY_j,\ZZ_{\D_j'^c})
 &=
 \oH(\ZZ_{\D_j'}|\CC_{\D_j},\YY_j,\ZZ_{\D_j'^c})
 \notag
 \\
 &\leq
 \oH(\ZZ_{\D_j'}|\CC_{\D_j},\YY_j)
 \notag
 \\
 &=
 0.
 \label{eq:HZgCYZ>=0}
\end{align}
From (\ref{eq:HZgCYZ>=0}) and
$\oH(\ZZ_{\D_j'}|\CC_{\D_j'},\YY_j,\ZZ_{\D_j'^c})\geq 0$,
we have
\begin{equation}
 \oH(\ZZ_{\D_j'}|\CC_{\D_j'},\YY_j,\ZZ_{\D_j'^c})=0
 \label{eq:HZgCYZ=0}
\end{equation}
for any $(j,\D_j')$ satisfying $j\in\J$ and $\emptyset\neq\D_j'\subset\D_j$.

Next, we show the relation
\begin{figure*}
\normalsize
\begin{align}
 &
 \oH(\CC_{\D_j'})+\oH(\ZZ_{\D_j'}|\CC_{\D_j'},\CC_{\D_j'^c},\YY_j,\ZZ_{\D_j'^c})
 \notag
 \\*
 &=
 \plimsupn\frac 1n\log_2\frac1{\mu_{C_{\D_j',n}}(C_{\D_j',n})}
 +
 \plimsupn\frac 1n\log_2
 \frac1{
  \mu_{Z_{\D_j'}^n|C_{\D_j',n}C_{\D_j'^c,n}Y_j^nZ_{\D_j'^c}^n}
  (Z_{\D_j'}^n|C_{\D_j',n},C_{\D_j'^c,n},Y_j^n,Z_{\D_j'^c}^n)
 }
 \notag
 \\
 &\geq
 \plimsupn\frac 1n\log_2
 \frac1{
  \mu_{C_{\D_j',n}}(C_{\D_j',n})
  \mu_{Z_{\D_j'}^n|C_{\D_j',n}C_{\D_j'^c,n}Y_j^nZ_{\D_j'^c}^n}
  (Z_{\D_j'}^n|C_{\D_j',n},C_{\D_j'^c,n},Y_j^n,Z_{\D_j'^c}^n)
 }
 \notag
 \\
 &=
 \plimsupn
 \frac 1n\log_2\frac
 {
  \mu_{C_{\D_j',n}|Y_j^nZ_{\D_j'^c}^n}(C_{\D_j',n}|Y_j^n,Z_{\D_j'^c}^n)
 }{
  \mu_{Z_{\D_j'}^n|Y_j^nZ_{\D_j'^c}^n}(Z_{\D_j'}^n|Y_j^n,Z_{\D_j'^c}^n)
  \mu_{C_{\D_j',n}|Z_{\D_j'}^n}(C_{\D_j',n}|Z_{\D_j'}^n)
  \mu_{C_{\D_j',n}}(C_{\D_j',n})
 }
 \notag
 \\
 &\geq
 \plimsupn
 \frac 1n\log_2\frac1
 {\mu_{Z_{\D_j'}^n|Y_j^nZ_{\D_j'^c}^n}(Z_{\D_j'}^n|Y_j^n,Z_{\D_j'^c}^n)}
 +
 \pliminfn
 \frac 1n\log_2\frac1
 {\mu_{C_{\D_j',n}|Z_{\D_j'}^n}(C_{\D_j',n}|Z_{\D_j'}^n)}
 \notag
 \\*
 &\quad
 +
 \pliminfn
 \frac 1n\log_2\frac
 {\mu_{C_{\D_j',n}|Y_j^nZ_{\D_j'^c}^n}(C_{\D_j',n}|Y_j^n,Z_{\D_j'^c}^n)}
 {\mu_{C_{\D_j',n}}(C_{\D_j',n})}
 \notag
 \\
 &=
 \oH(\ZZ_{\D_j'}|\YY_j,\ZZ_{\D_j'^c})
 +
 \uH(\CC_{\D_j'}|\ZZ_{\D_j'})
 +
 \uI(\CC_{\D_j'};\YY_j,\ZZ_{\D_j'^c})
 \label{eq:HUgVW+HV}
\end{align}
\hrulefill
\vspace*{4pt}
\end{figure*}
(\ref{eq:HUgVW+HV}), which appears on the top of the next page,
where the second equality comes from the fact that
$C_{\D_j',n}\markov Z_{\D_j'}^n\markov
(C_{\D_j'^c,n},Y_j^n,Z_{\D_j'^c}^n)$
and
$C_{\D_j'^c,n}\markov Z_{\D_j'^c}^n\markov
(C_{\D_j',n},Y_j^n,Z_{\D_j'}^n)$
form Markov chains, and inequalities comes from
(\ref{eq:plimsup-upper}),
(\ref{eq:plimsup-lower}), and (\ref{eq:pliminf-lower}).

Finally, we have
\begin{align}
 &
 \sum_{s\in\D_j'}r_s
 \notag
 \\*
 &\geq
 \sum_{s\in\D_j'}\limsupn \frac{\log_2|\C_{s,n}|}n
 \notag
 \\
 &\geq
 \limsupn \frac{\log_2|\Prod_{s\in\D_j'}\C_{s,n}|}n
 \notag
 \\
 &\geq
 \oH(\CC_{\D_j'})
 \notag
 \\
 &=
 \oH(\CC_{\D_j'})
 +\oH(\ZZ_{\D_j'}|\CC_{\D_j'},\CC_{\D_j'^c},\YY_j,\ZZ_{\D_j'^c})
 \notag
 \\
 &\geq
 \oH(\ZZ_{\D_j'}|\YY_j,\ZZ_{\D_j'^c})
 +\uH(\CC_{\D_j'}|\ZZ_{\D_j'})
 +\uI(\CC_{\D_j'};\YY_j,\ZZ_{\D_j'^c})
 \notag
 \\
 &\geq
 \oH(\ZZ_{\D_j'}|\YY_j,\ZZ_{\D_j'^c})
\end{align}
for all $(j,\D_j')$ satisfying $j\in\J$ and $\emptyset\neq\D_j'\subset\D_j$, where
the second inequality comes from the fact that
$C_{s,n}\in\M_{s,n}$,
the second equality comes from (\ref{eq:HZgCYZ=0}),
the third inequality comes from (\ref{eq:HUgVW+HV}),
and the last inequality comes from the fact that
$\uH(\CC_{\D_j'}|\ZZ_{\D_j'})\geq0$
and
$\uI(\CC_{\D_j'};\YY_j,\ZZ_{\D_j'^c})\geq 0$.
\hfill\IEEEQED

\subsection{Proof of $\ROP\subset\RIT$}
\label{sec:converse}

In the following, we prove $\ROP\subset\RIT$.

Assume that $\{R_s\}_{s\in\cS}\in\ROP$.
Then there is a code
$\{(\{\Phi_{i,n}\}_{i\in\I},\{\Psi_{j,n}\}_{j\in\J})\}_{n=1}^{\infty}$
that satisfies (\ref{eq:ROP-rate}) and (\ref{eq:ROP-error})
for all $i\in\I$ and $j\in\J$.

For $j\in\J$ and $\D_j'\subset\D_j$,
let
$\Psi_{j,\D_j',n}(Y^n_j)$
be the projection of $\Psi_{j,n}(Y^n_j)$ on $\Prod_{s\in\D_j'}\M_{s,n}$.
Then we have
\begin{equation*}
 \limn P(\Psi_{j,\D_j',n}(Y_j^n)\neq M_{\D_j',n})=0
\end{equation*}
from (\ref{eq:ROP-error}).
From Lemma~\ref{lem:fano}, we have
\begin{align}
 \oH(\MM_{\D_j'}|\YY_j,\MM_{\D_j'^c})
 &\leq
 \oH(\MM_{\D_j'}|\YY_j)
 \notag
 \\
 &=0.
 \label{eq:HMgYM>=0}
\end{align}
From (\ref{eq:HMgYM>=0})
and $\oH(\MM_{\D_j'}|\YY_j,\MM_{\D_j'^c})\geq 0$,
we have 
\begin{equation*}
 \oH(\MM_{\D_j'}|\YY_j,\MM_{\D_j'^c})=0
\end{equation*}
for any $(j,\D_j')$ satisfying $j\in\J$ and $\emptyset\neq\D_j'\subset\D_j$.
Let $r_s\equiv0$ for each $s\in\cS$.
Then it is clear that
\begin{align}
 \sum_{s\in\D_j'}r_s\geq \oH(\MM_{\D_j'}|\YY_j,\MM_{\D_j'^c})
 \label{eq:proof-converse-r}
\end{align}
for all $(j,\D_j')$ satisfying $j\in\J$ and $\emptyset\neq\D_j'\subset\D_j$.

Assume that $s\in\cS$.
Since the distribution $\mu_{M_{s,n}}$ of $M_{s,n}$ is uniform on
$\M_{s,n}$,
we have the fact that
\begin{align}
 \frac1n\log\frac 1{\mu_{M_{s,n}}(\mm_s)}
 &=
 \frac1n\log|\M_{s,n}|
 \notag
 \\
 &\geq
 \liminfn\frac 1n \log|\M_{s,n}|-\delta
\end{align}
for all $\mm_s\in\M_{s,n}$, $\delta>0$, and sufficiently large $n$.
This implies that
\begin{align}
 &\limn
 \Prob\lrsb{
  \frac1n\log\frac 1{\mu_{M_{s,n}}(M_{s,n})}
  <\liminfn\frac 1n \log|\M_{s,n}|-\delta
 }
 \notag
 \\*
 &=
 0,
 \label{eq:proof-converse-PM}
\end{align}
where we use the fact that $\mu_{M_{s,n}}(\mm_s)=0$
for all $\mm_s\notin\M_{s,n}$.
Let $\MM_s\equiv\{M_{s,n}\}_{n=1}^{\infty}$ be a general source\footnote{
 We can assume that $\M_{s,n}\subset\Z^n_s$ without loss of generality.}.
Then we have 
\begin{equation}
 \liminfn\frac 1n \log|\M_{s,n}|-\delta\leq \uH(\MM_s)
 \label{eq:proof-converse-uHM}
\end{equation}
from (\ref{eq:proof-converse-PM}) and the definition of $\uH(\MM_s)$.
We have
\begin{align}
 R_s+r_s
 &=
 R_s
 \notag
 \\
 &\leq
 \liminfn\frac{\log|\M_{s,n}|}n
 \notag
 \\
 &\leq
 \uH(\MM_s)+\delta
\end{align}
for all $s\in\cS$,
where the equality comes from the fact that $r_s=0$,
the first inequality comes from  (\ref{eq:ROP-rate}),
and the second inequality comes from  (\ref{eq:proof-converse-uHM}).
By letting $\delta\to0$, we have
\begin{align}
 R_s+r_s
 &\leq
 \uH(\MM_s).
 \label{eq:proof-converse-rR}
\end{align}
Let $\ZZ_s\equiv\MM_s$ for each $s\in\cS$
and $\XX_i\equiv\{\Phi_{i,n}(M_{\cS_i,n})\}_{n=1}^{\infty}$
for each $i\in\I$.
Since messages $\{M_s\}_{s\in\cS}$ are mutually independent,
the joint distribution of $(Z^n_{\cS},X^n_{\I},Y_{\J}^n)$ is given as
(\ref{eq:joint-rit}).
Then, from (\ref{eq:proof-converse-r}) and (\ref{eq:proof-converse-rR}),
we have $\{R_s\}_{s\in\cS}\in\RIT$, which implies $\ROP\subset\RIT$.
\hfill\IEEEQED

\subsection{$(\aalpha,\bbeta)$-hash property}
\label{sec:hash}

In this section, we review the hash property
introduced in \cite{CRNG}\cite{ISIT2010} and show two basic lemmas.
For the set $\F$ of functions,
let $\im\F\equiv \bigcup_{F\in\F}\{F\zz: \zz\in\Z^n\}$.

\begin{df}[{\cite[Definition~3]{CRNG}}]
Let $\F_n$ be a set of functions on $\U^n$.
For a probability distribution $p_{F_n}$ on $\F_n$, we
call a pair $(\F_n,p_{F_n})$ an {\em ensemble}.
Then, $(\F_n,p_{F_n})$ has an $(\alpha_{F_n},\beta_{F_n})$-{\em hash
 property} if
there is a pair $(\alpha_{F_n},\beta_{F_n})$
depending on $p_{F_n}$ such that
\begin{align}
 \sum_{\substack{
   \zz'\in\U^n\setminus\{\zz\}:
   \\
   p_{F_n}(\{f: f(\zz) = f(\zz')\})>\frac{\alpha_{F_n}}{|\im\F_n|}
 }}
 p_{F_n}\lrsb{\lrb{f: f(\zz) = f(\zz')}}
 \leq
 \beta_{F_n}
 \label{eq:hash}
\end{align}
for any $\zz\in\Z^n$.
Consider the following conditions for two sequences
$\aalpha_F\equiv\{\alpha_{F_n}\}_{n=1}^{\infty}$ and
$\bbeta_F\equiv\{\beta_{F_n}\}_{n=1}^{\infty}$
\begin{align}
 \limn \alpha_{F_n}
 &=1
 \label{eq:alpha-bcp}
 \\
 \limn \frac 1n\log(1+\beta_{F_n})
 &=0
 \label{eq:beta-bcp}
 \\
 \limn \frac 1n\log\alpha_{F_n}
 &=0
 \label{eq:alpha-crp}
 \\
 \limn \beta_{F_n}
 &=0.
 \label{eq:beta-crp}
\end{align}
Then, we say that 
$(\bcF,\bp_F)$ has an $(\aalpha_F,\bbeta_F)$-{\em balanced-coloring property}
if $\aalpha_F$ and $\bbeta_F$
satisfy (\ref{eq:hash}), (\ref{eq:alpha-bcp}), and (\ref{eq:beta-bcp}).
We say that 
$(\bcF,\bp_F)$ has an $(\aalpha_F,\bbeta_F)$-{\em collision-resistant property}
if $\aalpha_F$ and $\bbeta_F$
satisfy (\ref{eq:hash}), (\ref{eq:alpha-crp}), and (\ref{eq:beta-crp}).
We say that 
$(\bcF,\bp_F)$ has an $(\aalpha_F,\bbeta_F)$-{\em hash property}
if $\aalpha_F$ and $\bbeta_F$
satisfy (\ref{eq:hash}), (\ref{eq:alpha-bcp}), and (\ref{eq:beta-crp}).
Throughout this paper,
we omit the dependence of $\F$ and $F$ on $n$.
\end{df}

It should be noted that
when $\F$ is a two-universal class of hash functions \cite{CW}
and  $p_F$ is the uniform distribution on $\F$,
then $(\bcF,\bp_F)$ has a $(\one,\zero)$-hash property.
Random binning \cite{C75}
and the set of all linear functions \cite{CSI82} are
two-universal classes of hash functions.
It is proved in \cite[Section III-B]{HASH-BC} that
an ensemble of sparse matrices has a hash property.
It is proved in \cite[Section IV-B]{CRNGVLOSSY}
that an ensemble of systematic\footnote{The square part of the matrix is
 identity.}
sparse matrices has a balanced-coloring property.

We introduce lemmas that are multiple extensions of
the {\it balanced-coloring property} and
the {\it collision-resistant property}.
We use the following notations.
For each $s\in\cS$, let $\F_s$ be a set
of functions on $\Z_s^n$ and $\cc_s\in\im\F_s$.
Let $\Z_{\cS'}^n\equiv\Prod_{s\in\cS'}\Z_s^n$ and 
\begin{align*}
 \alpha_{F_{\cS'}}
 &\equiv
 \prod_{s\in\cS'}\alpha_{F_s}
 \\
 \beta_{F_{\cS'}}
 &\equiv
 \prod_{s\in\cS'}\lrB{\beta_{F_s}+1}-1,
\end{align*}
where $\prod_{s\in\emptyset}\theta_s\equiv1$.
It should be noted that
\begin{align*}
 \limn \alpha_{F_{\cS'}}=1
 \\
 \limn \frac1n\log(1+\beta_{F_{\cS'}})=0
 \\
 \limn \frac 1n\log\alpha_{F_{\cS'}}=0
 \\
 \limn \beta_{F_{\cS'}}=0
\end{align*}
for every $\cS'\subset\cS$
when $(\aalpha_{F_s},\bbeta_{F_s})$ satisfies
(\ref{eq:alpha-bcp}), (\ref{eq:beta-bcp}), (\ref{eq:alpha-crp}), and (\ref{eq:beta-crp}), respectively, for all $s\in\cS$.
For $\T\subset\Z_{\cS}^n$ and $\zz_{\cS'}\in\Z^n_{\cS'}$,
let $\T_{\cS'}$ and $\T_{\cS'^c|\cS'}(\zz_{\cS'})$ be defined as
\begin{align*}
 &
 \T_{\cS'}
 \equiv\{\zz_{\cS'}:
  (\zz_{\cS'},\zz_{\cS'^c})\in\T
  \ \text{for some}\ \zz_{\cS'^c}\in\Z_{\cS'^c}
  \}
 \\
 &
 \T_{\cS'^c|\cS'}(\zz_{\cS'})
 \equiv
 \{\zz_{\cS'^c}: (\zz_{\cS'},\zz_{\cS'^c})\in\T\}.
\end{align*}

The following lemma is related to the {\em balanced-coloring property},
which is an extension of the leftover hash lemma~\cite{IZ89}
and the balanced-coloring lemma~\cite[Lemma 3.1]{AC98}\cite[Lemma 17.3]{CK11}.
This lemma implies that there is an assignment that splits
a set equally.
\begin{lem}
\label{lem:mBCP}
For each $s\in\cS$, let $\F_s$ be a set
of functions on $\Z_s^n$
and $p_{F_s}$ be the probability distribution on $\F_s$,
where $(\F_s,p_{F_s})$ satisfies (\ref{eq:hash}).
We assume that the random variables
$F_{\cS}\equiv\{F_s\}_{s\in\cS}$
are mutually independent.
Then
\begin{align*}
 &
 E_{F_{\cS}}\lrB{
  \sum_{\cc_{\cS}}
  \lrbar{
   \frac{Q(\T\cap\fC_{F_{\cS}}(\cc_{\cS}))}
   {Q(\T)}
   -
   \frac1
   {\prod_{s\in\cS}|\im\F_s|}
  }
 }
 \notag
 \\*
 &\leq
 \sqrt{
  \alpha_{F_{\cS}}-1
  +
  \frac{
   \sum_{\substack{
     \cS'\subset\cS:
     \\
     \cS'\neq\emptyset
   }}
   \alpha_{F_{\cS'^c}}\!
   \lrB{\beta_{F_{\cS}}\!+\!1}\!
   \lrB{\prod_{s\in\cS'}\!|\im\F_s|}\!
   \bQ_{\cS'^c}
  }
  {Q(\T)}
 }
\end{align*}
for any function $Q:\Z_{\cS}\to[0,\infty)$ and $\T\subset\Z_{\cS}^n$,
where
\begin{equation}
 \bQ_{\cS'^c}
 \equiv
 \begin{cases}
  \displaystyle
  \max_{\zz_{\cS}\in\T}Q(\zz_{\cS})
  &\!\!\text{if}\ \cS'^c=\cS
  \\
  \displaystyle
  \max_{\zz_{\cS'}\in\T_{\cS'}}
  \!\!\!
  \sum_{\zz_{\cS'^c}\in\T_{\cS'^c|\cS'}(\zz_{\cS'})}
  \!\!\!
  Q(\zz_{\cS'},\zz_{\cS'^c})
  &\!\!\text{if}\ \emptyset\neq\cS'^c\subsetneq\cS
 \end{cases}
 \label{eq:maxQJ}
\end{equation}
\end{lem}
\begin{IEEEproof}
Let
$p_{\zz_s,\zz'_s}
\equiv
p_{F_s}\lrsb{\lrb{
  f_s:
  f_s(\zz_s)=f_s(\zz'_s)
}}$
and let $C_{\cS}$ be the random variable
corresponding to the uniform distribution on $\Prod_{s\in\cS}\im\F_s$.
In the following, we use the relation
\begin{align}
 \sum_{\substack{
   \zz_s\in\Z^n_s
   \\
   p_{\zz_s,\zz'_s}
   >\frac{\alpha_{F_s}}{|\im\F_s|}
 }}
 p_{\zz_s,\zz'_s}
 &=
 \sum_{\substack{
   \zz_s\in\Z^n_s\setminus\{\zz'_s\}
   \\
   p_{\zz_s,\zz'_s}
   >\frac{\alpha_{F_s}}{|\im\F_s|}
 }}
 p_{\zz_s,\zz'_s}
 +
 p_{\zz'_s,\zz'_s}
 \notag
 \\
 &\leq
 \beta_{F_s}+1
 \label{eq:proof-beta}
\end{align}
for all $\zz'_s\in\Z_s^n$, which comes from (\ref{eq:hash})
and the fact that $p_{\zz'_s,\zz'_s}=1$,

First, we have
\begin{align}
 &
 \sum_{\substack{
   \zz_{\cS}\in\T
   \\
   p_{\zz_s,\zz'_s}
   >\frac{\alpha_{F_s}}{|\im\F_s|}
   \ \text{for all}\  s\in\cS'
   \\
   p_{\zz_s,\zz'_s}
   \leq\frac{\alpha_{F_s}}{|\im\F_s|}
   \ \text{for all}\ s\in\cS'^c
 }}
 Q(\zz_{\cS})
 \prod_{s\in\cS} p_{\zz_s,\zz'_s}
 \notag
 \\*
 &=
 \sum_{\substack{
   \zz_{\cS'}\in\T_{\cS'}
   \\
   p_{\zz_s,\zz'_s}
   >\frac{\alpha_{F_s}}{|\im\F_s|}
 }}
 \lrB{\prod_{s\in\cS'} p_{\zz_s,\zz'_s}}
 \notag
 \\*
 &\quad\cdot
 \sum_{\substack{
   \zz_{\cS'^c}\in\T_{\cS'^c|\cS'}\lrsb{\zz_{\cS'}}:
   \\
   p_{\zz_s,\zz'_s}
   \leq\frac{\alpha_{F_s}}{|\im\F_s|}
 }}
 Q(\zz_{\cS'},\zz_{\cS'^c})
 \prod_{s\in\cS'^c}p_{\zz_s,\zz'_s}
 \notag
 \\
 &\leq
 \lrB{\prod_{s\in\cS'^c}\frac{\alpha_{F_s}}{|\im\F_s|}}
 \sum_{\substack{
   \zz_{\cS'}\in\T_{\cS'}:
   \\
   p_{\zz_s,\zz'_s}
   >\frac{\alpha_{F_s}}{|\im\F_s|}
 }}
 \lrB{
  \prod_{s\in\cS'} p_{\zz_s,\zz'_s}
 }
 \notag
 \\*
 &\quad\cdot
 \sum_{\zz_{\cS'^c}\in\T_{\cS'^c|\cS'}\lrsb{\zz_{\cS'}}}
 Q(\zz_{\cS'},\zz_{\cS'^c})
 \notag
 \\
 &\leq
 \bQ_{\cS'^c}
 \lrB{\prod_{s\in\cS'^c}\frac{\alpha_{F_s}}{|\im\F_s|}}
 \prod_{s\in\cS'}
 \lrB{
  \sum_{\substack{
    \zz_s\in\Z_s^n:
    \\
    p_{\zz_s,\zz'_s}
    >\frac{\alpha_{F_s}}{|\im\F_s|}
  }}
  p_{\zz_s,\zz'_s}
 }
 \notag
 \\
 &\leq
 \bQ_{\cS'^c}
 \lrB{\prod_{s\in\cS'^c}\frac{\alpha_{F_s}}{|\im\F_s|}}
 \prod_{s\in\cS'}
 \lrB{\beta_{F_s}+1}
 \notag
 \\
 &=
 \frac{
  \alpha_{F_{\cS'^c}}\lrB{\beta_{F_{\cS'}}+1}
  \bQ_{\cS'^c}
 }
 {\prod_{s\in\cS'^c}\lrbar{\im\F_s}}
 \label{eq:lemma-multi-subset}
\end{align}
for all $(\zz'_{\cS},\cS')$ satisfying
$\zz'_{\cS}\in\T$ and $\emptyset\neq\cS'\subsetneq\cS$,
where
the second inequality comes from (\ref{eq:maxQJ})
and the third inequality comes from (\ref{eq:proof-beta}).
It should be noted that (\ref{eq:lemma-multi-subset})
is valid for the cases $\cS'^c=\emptyset$ and $\cS'^c=\cS$
by letting $\bQ_{\emptyset}\equiv Q(\T)$
because
\begin{align}
 \sum_{\substack{
   \zz_{\cS}\in\T:
   \\
   p_{\zz_s,\zz'_s}
   \leq\frac{\alpha_{F_s}}{|\im\F_s|}
   \ \text{for all}\ s\in\cS
 }}
 Q(\zz_{\cS})
 \prod_{s\in\cS} p_{\zz_s,\zz'_s}
 &\leq
 \frac{\alpha_{F_{\cS}}Q(\T)}
 {\prod_{s\in\cS}\lrbar{\im\F_s}}
 \notag
 \\
 &=
 \frac{\alpha_{F_{\cS}}\lrB{\beta_{F_{\emptyset}}+1}\bQ_{\emptyset}}
 {\prod_{s\in\cS}\lrbar{\im\F_s}}
\end{align}
and
\begin{align}
 &
 \sum_{\substack{
   \zz_{\cS}\in\T:
   \\
   p_{\zz_s,\zz'_s}
   >\frac{\alpha_{F_s}}{|\im\F_s|}
   \ \text{for all}\ s\in\cS
 }}
 Q(\zz_{\cS})
 \prod_{s\in\cS} p_{\zz_s,\zz'_s}
 \notag
 \\*
 &\leq
 \lrB{\max_{\zz_{\cS}\in\T}Q(\zz_{\cS})}
 \sum_{\substack{
   \zz_{\cS}\in\T:
   \\
   p_{\zz_s,\zz'_s}
   >\frac{\alpha_{F_s}}{|\im\F_s|}
   \ \text{for all}\ s\in\cS
 }}
 \prod_{s\in\cS} p_{\zz_s,\zz'_s}
 \notag
 \\
 &\leq
 \lrB{\max_{\zz_{\cS}\in\T}Q(\zz_{\cS})}
 \prod_{s\in\cS}
 \lrB{
  \sum_{\substack{
    \zz_s\in\Z_s^n:
    \\
    p_{\zz_s,\zz'_s}
    >\frac{\alpha_{F_s}}{|\im\F_s|}
  }}
  p_{\zz_s,\zz'_s}
 }
 \notag
 \\
 &\leq
 \lrB{\max_{\zz_{\cS}\in\T}Q(\zz_{\cS})}
 \prod_{s\in\cS}
 \lrB{\beta_{F_s}+1}
 \notag
 \\
 &=
 \frac{\alpha_{F_{\emptyset}}\lrB{\beta_{F_{\cS}}+1}\bQ_{\cS}}
 {\prod_{s\in\emptyset}\lrbar{\im\F_s}}.
 \label{eq:proof-BCP-multi1}
\end{align}
Then we have
\begin{align}
 &
 \sum_{\zz_{\cS}\in\T}Q(\zz_{\cS})
 \prod_{s\in\cS}p_{\zz_s,\zz'_s}
 \notag
 \\*
 &\leq
 \sum_{\cS'\subset\cS}
 \sum_{\substack{
   \zz_{\cS}\in\T:
   \\
   p_{\zz_s,\zz'_s}
   >\frac{\alpha_{F_s}}{|\im\F_s|}
   \ \text{for all}\ s\in\cS'
   \\
   p_{\zz_s,\zz'_s}
   \leq\frac{\alpha_{F_s}}{|\im\F_s|}
   \ \text{for all}\ s\in\cS'^c
 }}
 Q(\zz_{\cS})
 \prod_{s\in\cS} p_{\zz_s,\zz'_s}
 \notag
 \\
 &\leq
 \sum_{\cS'\subset\cS}
 \frac{
  \alpha_{F_{\cS'^c}}\lrB{\beta_{F_{\cS'}}+1}
  \bQ_{\cS'^c}
 }{\prod_{s\in\cS'^c}\lrbar{\im\F_s}}
 \notag
 \\
 &
 =
 \frac{\alpha_{F_{\cS}}Q(\T)}
 {\prod_{s\in\cS}\lrbar{\im\F_s}}
 +
 \sum_{\substack{
   \cS'\subset\cS:
   \\
   \cS'\neq\emptyset
 }}
 \frac{
  \alpha_{F_{\cS'^c}}\lrB{\beta_{F_{\cS'}}+1}
  \bQ_{\cS'}
 }
 {\prod_{s\in\cS'^c}\lrbar{\im\F_s}}
 \label{eq:proof-BCP-multi2}
\end{align}
for all $\zz'_{\cS}\in\T$,
where the equality comes from the fact that $\bQ_{\emptyset}=Q(\T)$ and
$\beta_{F_{\emptyset}}=0$.

Next, let $C_{\cS}$ be the random variable subject to the uniform
distribution on $\Prod_{s\in\cS}\im\F_s$.
From (\ref{eq:proof-BCP-multi2}), we have
\begin{align}
 &
 E_{F_{\cS}C_{\cS}}\lrB{
  \lrB{
   \sum_{\zz_{\cS}\in\T}Q(\zz_{\cS})\chi(F_{\cS}(\zz_{\cS})=C_{\cS})
  }^2
 }
 \notag
 \\*
 &=
 \sum_{\zz'_{\cS}\in\T}Q(\zz'_{\cS})
 \sum_{\zz_{\cS}\in\T}Q(\zz_{\cS})
 \notag
 \\*
 &\quad
 \cdot
 E_{F_{\cS}}\lrB{
  \chi(F_{\cS}(\zz_{\cS})=F_{\cS}(\zz'_{\cS}))
  E_{C_{\cS}}\lrB{\chi(F_{\cS}(\zz_{\cS})=C_{\cS})}
 }
 \notag
 \\
 &=
 \frac1{\prod_{s\in\cS}\lrbar{\im\F_s}}
 \sum_{\zz'_{\cS}\in\T}Q(\zz'_{\cS})
 \sum_{\zz_{\cS}\in\T}Q(\zz_{\cS})
 \prod_{s\in\cS}p_{\zz_s,\zz'_s}
 \notag
 \\
 &\leq
 \frac{\alpha_{F_{\cS}}Q(\T)^2}
 {\lrB{\prod_{s\in\cS}\lrbar{\im\F_s}}^2}
 \notag
 \\*
 &\quad
 +
 \frac{Q(\T)}
 {\prod_{s\in\cS}\lrbar{\im\F_s}}
 \sum_{\substack{
   \cS'\subset\cS:
   \\
   \cS'\neq\emptyset
 }}
 \frac{
  \alpha_{F_{\cS'^c}}\lrB{\beta_{F_{\cS'}}+1}
  {\bQ_{\cS'^c}}
 }
 {\prod_{s\in\cS'^c}\lrbar{\im\F_s}}.
 \label{eq:lemma-multi}
\end{align}
Then we have
\begin{align}
 &
 E_{F_{\cS}C_{\cS}}\lrB{
  \lrB{
   \frac{Q\lrsb{\T\cap\fC_{F_{\cS}}(C_{\cS})}
    \prod_{s\in\cS}|\im\F_s|}
   {Q(\T)}
   -1}^2
 }
 \notag
 \\*
 &=
 E_{F_{\cS}C_{\cS}}\lrB{
  \lrB{\sum_{\zz_{\cS}\in\T}
   \frac{Q(\zz)\chi(F_{\cS}(\zz_{\cS})=C_{\cS})
    \prod_{s\in\cS}|\im\F_s|
   }{Q(\T)}
  }^2
 }
 \notag
 \\*
 &\quad
 -2
 E_{F_{\cS}C_{\cS}}\lrB{
  \sum_{\zz_{\cS}\in\T}
  \frac{Q(\zz)\chi(F_{\cS}(\zz_{\cS})=C_{\cS})
   \prod_{s\in\cS}|\im\F_s|
  }{Q(\T)}
 }
 \notag
 \\*
 &\quad
 +1
 \notag
 \\
 &=
 E_{F_{\cS}C_{\cS}}\lrB{
  \lrB{\sum_{\zz_{\cS}\in\T}
   \frac{Q(\zz)\chi(F_{\cS}(\zz_{\cS})=C_{\cS})
    \prod_{s\in\cS}|\im\F_s|
   }{Q(\T)}
  }^2
 }
 \notag
 \\*
 &\quad
 -2
 \sum_{\zz_{\cS}\in\T}
 \frac{
  Q(\zz)
  E_{F_{\cS}C_{\cS}}\lrB{\chi(F_{\cS}(\zz_{\cS})=C_{\cS})}
  \prod_{s\in\cS}|\im\F_s|
 }{Q(\T)}
 \notag
 \\*
 &\quad
 +1
 \notag
 \\
 &=
 \frac{\displaystyle
  \lrB{\prod_{s\in\cS}|\im\F_s|}^2
 }{Q(\T)^2}
 E_{F_{\cS}C_{\cS}}\lrB{
  \lrB{\sum_{\zz_{\cS}\in\T}
   Q(\zz_{\cS})\chi(F_{\cS}(\zz_{\cS})=C_{\cS})
  }^2
 }
 \notag
 \\*
 &\quad
 -1
 \notag
 \\
 &\leq
 \alpha_{F_{\cS}}-1
 +
 \frac{
  \sum_{\substack{
    \cS'\subset\cS
    \\
    \cS'\neq\emptyset
  }}
  \alpha_{F_{\cS'^c}}\lrB{\beta_{F_{\cS'}}+1}
  \lrB{\prod_{s\in\cS'}\lrbar{\im\F_s}}\bQ_{\cS'^c}
 }
 {Q(\T)},
\end{align}
where
the inequality comes from 
(\ref{eq:lemma-multi}).

Finally, the lemma is shown as
\begin{align}
 &
 E_{F_{\cS}}\lrB{
  \sum_{\cc_{\cS}}
  \left|
   \frac{Q\lrsb{\T\cap\fC_{F_{\cS}}(\cc_{\cS})}}{Q(\T)}
   -\frac 1{\prod_{s\in\cS}|\im\F_s|}
  \right|
 }
 \notag
 \\*
 &=
 E_{F_{\cS}C_{\cS}}\lrB{
  \left|
   \frac{Q\lrsb{\T\cap\fC_{F_{\cS}}(C_{\cS})}
    \prod_{s\in\cS}|\im\F_s|
   }{Q(\T)}
   -1
  \right|
 }
 \notag
 \\
 &=
 E_{F_{\cS}C_{\cS}}\lrB{
  \sqrt{
   \lrB{
    \frac{
     Q\lrsb{\T\cap\fC_{F_{\cS}}(C_{\cS})}
     \prod_{s\in\cS}|\im\F_s|
    }{Q(\T)}
    -1}^2
  }
 }
 \notag
 \\
 &\leq
 \sqrt{
  E_{F_{\cS}C_{\cS}}\lrB{
   \lrB{\frac{
     Q\lrsb{\T\cap\fC_{F_{\cS}}(C_{\cS})}
     \prod_{s\in\cS}|\im\F_s|
    }{Q(\T)}
    -1}^2
  }
 }
 \notag
 \\
 &\leq
 \sqrt{
  \alpha_{F_{\cS}}-1
  +
  \frac{
   \sum_{\substack{
     \cS'\subset\cS:
     \\
     \cS'\neq\emptyset
   }}
   \alpha_{F_{\cS'^c}}\!
   \lrB{\beta_{F_{\cS}}\!+\!1}\!
   \lrB{\prod_{s\in\cS'}\!|\im\F_s|}\!
   \bQ_{\cS'^c}
  }
  {Q(\T)}
 },
 \label{eq:proof-BCP-multi}
\end{align}
where the first inequality comes from the Jensen inequality.
\end{IEEEproof}

The following lemma is a multiple extension of
the {\it collision-resistant property}.
This lemma implies that
there is an assignment such that every bin contains at most one
item.
\begin{lem}[{\cite[Lemma 7]{HASH-BC}}]
\label{lem:mCRP}
For each $s\in\cS$, let $\F_s$ be a set
of functions on $\Z_s^n$
and $p_{F_s}$ be the probability distribution on $\F_s$,
where $(\F_s,p_{F_s})$ satisfies (\ref{eq:hash}).
We assume that random variables
$F_{\cS}\equiv\{F_s\}_{s\in\cS}$
are mutually independent.
Then
\begin{align*}
 &
 p_{F_{\cS}}\lrsb{\lrb{
   f_{\cS}:
   \lrB{\T\setminus\{\zz_{\cS}\}}\cap\fC_{f_{\cS}}(f_{\cS}(\zz_{\cS}))
   \neq
   \emptyset
 }}
 \notag
 \\*
 &\quad
 \leq
 \sum_{\substack{
   \cS'\subset\cS:
   \\
   \cS'\neq\emptyset
 }}
 \frac{
  \alpha_{F_{\cS'}}\lrB{\beta_{F_{\cS'^c}}+1}
  \oO_{\cS'}
 }
 {
  \prod_{s\in\cS'}\lrbar{\im\F_s}
 }
 +\beta_{F_{\cS}}
\end{align*}
for all $\T\subset\Z_{\cS}^n$ and $\zz_{\cS}\in\Z_{\cS}^n$,
where
\begin{align*}
 \oO_{\cS'}
 &\equiv
 \begin{cases}
  |\T|
  &\text{if}\ \cS'=\cS,
  \\
  \displaystyle\max_{\zz_{\cS'^c}\in\T_{\cS'^c}}
  \lrbar{\T_{\cS'|\cS'^c}\lrsb{\zz_{\cS'^c}}},
  &\text{if}\ \emptyset\neq\cS'\subsetneq\cS
 \end{cases}
\end{align*}
\end{lem}
\begin{IEEEproof}
Let
$p_{\zz_s,\zz'_s}
\equiv
p_{F_s}\lrsb{\lrb{
  f_s:
  f_s(\zz_s)=f_s(\zz'_s)
}}$.
By interchanging $\cS'$ and $\cS'^c$,
and letting
$\oO_{\emptyset}=1$ and
$Q(\zz_{\cS})\equiv 1$ for each $\zz_{\cS}\in\Z_{\cS}^n$,
we have the fact that
\begin{align}
 \sum_{\substack{
   \zz_{\cS'}\in\T
   \\
   p_{\zz_s,\zz'_s}
   \leq\frac{\alpha_{F_s}}{|\im\F_s|}
   \ \text{for all}\ s\in\cS'
   \\
   p_{\zz_s,\zz'_s}
   >\frac{\alpha_{F_s}}{|\im\F_s|}
   \ \text{for all}\ s\in\cS'^c
 }}
 \prod_{s\in\cS'} p_{\zz_s,\zz'_s}
 &\leq
 \frac{\alpha_{F_{\cS'}}\lrB{\beta_{F_{\cS'^c}}+1}\oO_{\cS'}}
 {\prod_{s\in\cS'}|\im\F_s|}
 \label{eq:lemma-mcrp}
\end{align}
for all $\zz'_{\cS'}\in\Z_{\cS'}^n$ and $\cS'\subset\cS$ from
(\ref{eq:lemma-multi-subset}).
Then we have
\begin{align}
 &
 p_{F_{\cS}}\lrsb{\lrb{
   f_{\cS}:
   \lrB{\T\setminus\{\zz_{\cS}\}}\cap\fC_{F_{\cS}}(F_{\cS}\zz_{\cS})
   \neq
   \emptyset
 }}
 \notag
 \\*
 &\leq
 \sum_{\zz_{cS}\in\T\setminus\{\zz'_{\cS}\}}
 p_{F_{\cS}}\lrsb{\lrb{
   f_{\cS}:
   f_{\cS}(\zz_{\cS})=f_{\cS}(\zz'_{\cS})
 }}
 \notag
 \\
 &=
 \sum_{\zz_{\cS}\in\T\setminus\{\zz'_{\cS}\}}
 p_{F_{\cS}}\lrsb{\lrb{
   f_{\cS}:
   f_s(\zz_s)=f_s(\zz'_s)
   \ \text{for all}\ s\in\cS
 }}
 \notag
 \\*
 &=
 \sum_{\zz_{\cS}\in\T}
 \prod_{s\in\cS} p_{\zz_s,\zz'_s}
 -
 \prod_{s\in\cS} p_{\zz'_s,\zz'_s}
 \notag
 \\
 &=
 \sum_{\cS'\subset\cS}
 \sum_{\substack{
   \zz_{\cS}\in\T\setminus\{\zz'_{\cS}\}
   \\
   p_{\zz_s,\zz'_s}
   \leq\frac{\alpha_{F_s}}{|\im\F_s|}
   \ \text{for all}\ s\in\cS'
   \\
   p_{\zz_s,\zz'_s}
   >\frac{\alpha_{F_s}}{|\im\F_s|}
   \ \text{for all}\ s\in\cS'^c
 }}
 \prod_{s\in\cS} p_{\zz_s,\zz'_s}
 -1
 \notag
 \\
 &\leq
 \sum_{\cS'\subset\cS}
 \frac{\alpha_{F_{\cS'}}\lrB{\beta_{F_{\cS'^c}}+1}\oO_{\cS'}}
 {\prod_{s\in\cS}|\im\F_s|}
 -1
 \notag
 \\
 &=
 \sum_{\substack{
   \cS'\subset\cS:
   \\
   \cS'\neq\emptyset
 }}
 \frac{
  \alpha_{F_{\cS'}}\lrB{\beta_{F_{\cS'^c}}+1}\oO_{\cS'}
 }
 {\prod_{s\in\cS'}|\im\F_s|}
 +
 \beta_{F_{\cS}}
\end{align}
for all $\T\subset\Z_{\cS'}^n$ and $\zz'_{\cS'}\in\Z_{\cS'}^n$,
where the third equality comes from the fact that $p_{\zz'_s,\zz'_s}=1$,
the second inequality comes from (\ref{eq:lemma-mcrp}),
and the last equality comes from the fact that
$\alpha_{F_{\emptyset}}=1$,
$\beta_{F_{\emptyset^c}}=\beta_{F_{\cS}}$,
$\prod_{s\in\emptyset}|\im\F_s|=1$,
and $\oO_{\emptyset}=1$.
\end{IEEEproof}

\subsection{Proof of Theorem~\ref{thm:source}}
\label{sec:proof-source}

In the following,
we omit the dependence of
$Z$, $C$, $Y$, and $\hZ$ on $n$ when they appear in the subscript of $\mu$.

First, we prove the following lemma,
where we omit the dependence of $\YY$, $\D$, and $\delta$ on $j\in\J$.
\begin{lem}
\label{lem:source}
Let $(\ZZ_{\D},\YY)$ be a pair of general correlated sources
and
\begin{align*}
 \oT_{Z_{\D}}
 &\equiv
 \lrb{
  (\zz_{\D},\yy):
  \begin{aligned}
   &
   \frac1n\log
   \frac1{\mu_{Z_{\D'}|Z_{\D'^c}Y}
    (\zz_{\D'}|\zz_{\D'^c},\yy)}
   \\
   &\quad
   \leq
   \oH(\ZZ_{\D'}|\YY,\ZZ_{\D'^c})+\e
   \\
   &\text{for all}\ \D'\ \text{satisfying}\ \emptyset\neq\D'\subset\D
  \end{aligned}
 }.
\end{align*}
For a given $\{r_s\}_{s\in\D}$ satisfying (\ref{eq:rate-r})
for every $\D'$ satisfying $\emptyset\neq\D'\subset\D$,
assume that $(\bcF_s,\bp_{F_s})$
has the collision-resistant property for every $s\in\D_j$,
where $r_s=\log(|\im\F_s|)/n$.
Then for any $\delta>0$
and all sufficiently large $n$
there are functions (sparse matrices)
$f_{\D}\equiv\{f_s\}_{s\in\D}$ such that
\begin{equation*}
 \mu_{Z_{\D}^nY^n}\lrsb{\lrb{
   (\zz_{\D},\yy): \hzz_{\D}(f_{\D}(\zz_{\D})|\yy)\neq\zz_{\D}
 }}\leq \delta,
\end{equation*}
where $\hzz_{\D}(\cc_{\D}|\yy)$ outputs one of the elements
in $\oT_{Z_{\D}}\cap\fC_{f_{\D}}(\cc_{\D})$
and declares an error when $\oT_{Z_{\D}}\cap\fC_{f_{\D}}(\cc_{\D})=\emptyset$.
\end{lem}
\begin{IEEEproof}
Let $\oT_{Z_{\D}}(\yy)\equiv\{\zz_{\D}: (\zz_{\D},\yy)\in\oT_{Z_{\D}}\}$
and assume that $(\zz_{\D},\yy)\in\oT_{Z_{\D}}$
and $\hzz_{\D}(f_{\D}(\zz_{\D})|\yy)\neq\zz_{\D}$.
Then 
$\lrB{\oT_{Z_{\D}}(\yy)\setminus\{\zz_{\D}\}}
\cap\fC_{f_{\D}}(f_{\D}(\zz_{\D}))\neq\emptyset$.
We have
\begin{align}
 &
 E_{F_{\D}}\lrB{
  \chi(\hzz_{\D}(F_{\D}(\zz_{\D})|\yy)\neq\zz_{\D})
 }
 \notag
 \\*
 &\leq
 p_{F_{\D}}\lrsb{\lrb{
   f:
   \lrB{\oT_{\Z_{\D}}(\yy)\setminus\{\zz_{\D}\}}
   \cap\C_{f_{\D}}(f_{\D}(\zz_{\D}))\neq\emptyset
 }}
 \notag
 \\
 &\leq
 \sum_{\substack{
   \D'\subset\D:
   \\
   \D'\neq\emptyset
 }}
 \frac{
  \alpha_{F_{\D'}}\lrB{\beta_{F_{\D'^c}}+1}
  \oO_{\D'}
 }
 {
  \prod_{s\in\D'}\lrbar{\im\F_s}
 }
 +\beta_{F_{\D}}
 \notag
 \\
 \begin{split}
  &\leq
  \sum_{\substack{
    \D'\subset\D:\\
    \D'\neq\emptyset
  }}
  \alpha_{F_{\D'}}\lrB{\beta_{F_{\D'^c}}+1}
  2^{
   -n\lrB{\sum_{s\in\D'}r_s-\oH(\ZZ_{\D'}|\YY,\ZZ_{\D'^c})-\e}
  }
  \\*
  &\quad
  +\beta_{F_{\D}},
 \end{split}
\end{align}
where the second inequality comes from Lemma~\ref{lem:mCRP} and
the third inequality comes from 
$\oO_{\D'}\leq2^{n\lrB{\oH(\ZZ_{\D'}|\YY,\ZZ_{\D'^c})+\e}}$.
We have
\begin{align}
 &
 E_{F_{\D}}\lrB{\lrb{
   (\zz_{\D},\yy): \hzz_{\D}(F_{\D}(\zz_{\D})|\yy)\neq\zz_{\D}
 }}
 \notag
 \\*
 &=
 E_{F_{\D}}\lrB{
  \sum_{\zz_{\D},\yy}
  \mu_{Z_{\D}Y}(\zz_{\D},\yy)
  \chi(\hzz_{\D}(F_{\D}(\zz_{\D})|\yy)\neq \zz_{\D})
 }
 \notag
 \\*
 &=
 \sum_{(\zz_{\D},\yy)\in\oT_{Z_{\D}}}
 \mu_{Z_{\D}Y}(\zz_{\D},\yy)
 E_{F_{\D}}\lrB{
  \chi(\hzz_{\D}(F_{\D}(\zz_{\D})|\yy)\neq\zz_{\D})
 }
 \notag
 \\*
 &\quad
 +
 \!\!
 \sum_{(\zz_{\D},\yy)\in\oT_{Z_{\D}}^c}
 \!\!
 \mu_{Z_{\D}Y}(\zz_{\D},\yy)
 E_{F_{\D}}\lrB{
  \chi(\hzz_{\D}(F_{\D}(\zz_{\D})|\yy)\neq \zz_{\D})
 }
 \notag
 \\
 \begin{split}
  &\leq
  \sum_{\substack{
    \D'\subset\D:\\
    \D'\neq\emptyset
  }}
  \alpha_{F_{\D'}}\lrB{\beta_{F_{\D'^c}}+1}
  2^{
   -n\lrB{\sum_{s\in\D'}r_s-\oH(\ZZ_{\D'}|\YY,\ZZ_{\D'^c})-\e}
  }
  \\*
  &\quad
  +\beta_{F_{\D}}
  +\mu_{Z_{\D}Y}(\oT_{Z_{\D}}^c).
 \end{split}
\end{align}
From this inequality and the fact that
$\log(\alpha_{F_{\D'}})/n\to0$, $\beta_{F_{\D'}}\to0$,
$\mu_{Z_{\D}Y}(\oT_{Z_{\D}}^c)\to0$,
we have the fact that
for all $\delta>0$ and sufficiently large $n$
there are $\{f_s\}_{s\in\D}$
such that the error probability is less than $\delta$
for all sufficiently large $n$ 
when $\{r_s\}_{s\in\D}$
satisfies 
\begin{align*}
 \sum_{s\in\D'}r_s
 &>
 \oH(\ZZ_{\D'}|\ZZ_{\D'^c},\YY)+\e
\end{align*}
for all $\D'$ satisfying $\emptyset\neq\D'\subset\D$
by letting sufficiently small $\e>0$.
\end{IEEEproof}

Next, we introduce the following lemma.
\begin{lem}[{\cite[Corollary 2]{SDECODING}}]
\label{lem:sdecoding}
Let $(U,V)$ be a pair consisting of a state $U$ and an observation $V$
and $\mu_{UV}$ be the joint distribution of $(U,V)$.
We make a stochastic decision with $\mu_{U|V}$,
that is, the joint distribution of $(U,V)$ and a guess $\hU$ of the
state $U$ is given as
\begin{equation*}
 \mu_{UV\hU}(u,v,\hu)\equiv \mu_{UV}(u,v)\mu_{U|V}(\hu|v).
\end{equation*}
Then the decision error probability of this rule
is at most twice the decision error probability
of {\it any} (possibly stochastic) decision, 
that is,
\begin{equation*}
 \Prob\lrsb{\hU\neq U}
 \leq
 2\Prob\lrsb{\hU'\neq U},
\end{equation*}
where
\begin{equation*}
 \mu_{UV\hU'}(u,v,\hu)\equiv \mu_{UV}(u,v)\mu_{\hU'|V}(\hu|v)
\end{equation*}
for any probability distribution $\mu_{\hU'|V}$.
\end{lem}

Finally, we prove Theorem~\ref{thm:source}.
The joint distribution of 
$(Z_{\D_j}^n,C_{\D_j,n},Y_j^n)$ is given as
\begin{align*}
 &
 \mu_{Z_{\D_j}C_{\D_j}Y_j}(\zz_{\D_j},\cc_{\D_j},\yy_j)
 \notag
 \\*
 &=
 \mu_{Z_{\D_j}Y_j}(\zz_{\D_j},\yy_j)
 \chi(f_{\D_j}(\zz_{\D_j})=\cc_{\D_j}).
\end{align*}
Then we have
\begin{align}
 &
 \mu_{Z_{\D_j}|C_{\D_j}Y_j}(\zz_{\D_j}|\cc_{\D_j},\yy_j)
 \notag
 \\*
 &=
 \frac{\mu_{Z_{\D_j}C_{\D_j}Y_j}(\zz_{\D_j},\cc_{\D_j},\yy_j)}
 {\sum_{\zz_{\D_j}}\mu_{Z_{\D_j}C_{\D_j}Y_j}(\zz_{\D_j},\cc_{\D_j},\yy_j)}
 \notag
 \\
 &=
 \frac{\mu_{Z_{\D_j},Y_j}(\zz_{\D_j},\yy_j)\chi(f_{\D_j}(\zz_{\D_j})=\cc_{\D_j})}
 {\sum_{\zz_{\D_j}}\mu_{Z_{\D_j},Y_j}(\zz_{\D_j},\yy_j)
  \chi(f_{\D_j}(\zz_{\D_j})=\cc_{\D_j})}
 \notag
 \\
 &=
 \frac{\mu_{Z_{\D_j}|Y_j}(\zz_{\D_j}|\yy_j)\chi(f_{\D_j}(\zz_{\D_j})=\cc_{\D_j})}
 {\sum_{\zz_{\D_j}}\mu_{Z_{\D_j}|Y_j}(\zz_{\D_j}|\yy_j)
  \chi(f_{\D_j}(\zz_{\D_j})=\cc_{\D_j})}
 \notag
 \\
 &=
 \mu_{\hZ_{\D_j}|C_{\D_j}Y_j}(\zz_{\D_j}|\cc_{\D_j},\yy_j),
\end{align}
that is, the constrained-random-number generator defined by
(\ref{eq:source-decoder}) is a stochastic decision with
$\mu_{Z_{\D_j}|C_{\D_j}Y_j}$.
By applying Lemmas~\ref{lem:source} and~\ref{lem:sdecoding}, we have the fact that
\begin{align}
 &
 \Prob\lrsb{\hZ_{\D_j}^n\neq Z^n_{\D_j}}
 \notag
 \\*
 &\leq
 2\Prob\lrsb{
  \hzz_{\D_j}(f_{\D_j}(Z^n_{\D_j})|Y^n_j)
  \neq Z^n_{\D_j}
 }
 \notag
 \\
 &=
 2\mu_{Z_{\D_j}Y_j}\lrsb{\lrb{
   (\zz_{\D_j},\yy_j): \hzz_{\D_j}(f_{\D_j}(\zz_{\D_j})|\yy_j)\neq\zz_{\D_j}
 }}
 \notag
 \\
 &\leq
 2\delta_j
\end{align}
for given positive numbers $\{\delta_j\}_{j\in\J}$ and all sufficiently
large $n$.
When $\hZ_{j,s}^n\neq Z^n_s$ for some $j\in\J$ and $s\in\D_j$,
we have $\hZ_{\D_j}^n\neq Z^n_{\D_j}$ for some $j\in\J$.
From this fact, we have
\begin{align}
 E_{F_{\cS}}\lrB{
  \Error(F_{\cS})
 }
 &\leq
 E_{F_{\cS}}\lrB{
  \sum_{j\in\J}
  \Prob\lrsb{\hZ_{\D_j}^n\neq Z^n_{\D_j}}
 }
 \notag
 \\
 &=
 \sum_{j\in\J}
 E_{F_{\D_j}}\lrB{
  \Prob\lrsb{\hZ_{\D_j}^n\neq Z^n_{\D_j}}
 }
 \notag
 \\
 &\leq
 2\sum_{j\in\J}
 \delta_j
\end{align}
for all positive values $\{\delta_j\}_{j\in\J}$ and all sufficiently
large $n$.
We obtain the theorem by letting $2\sum_{j\in\J}\delta_j<\delta$.
\hfill\IEEEQED

\begin{figure*}[!t]
\normalsize
\begin{align}
 &
 \Error(f_{\cS},g_{\cS},\cc_{\cS})
 \notag
 \\*
 &=
 \sum_{\substack{
   \mm_{\cS}:\\
   \mu_{Z_{\cS_i}}(\fC_{f_{\cS_i}g_{\cS_i}}(\cc_{\cS_i},\mm_{\cS_i}))=0
   \ \text{for some}\ i\in\I
 }}
 \prod_{i\in\I}
 \frac1{\prod_{s\in\cS_i}|\im\G_s|}
 \notag
 \\*
 &\quad\quad
 +
 \sum_{\substack{
   \mm_{\cS},\zz_{\cS},\xx_{\I},\yy_{\J},\hzz_{\D_{\J}}:\\
   \mu_{Z_{\cS_i}}(\fC_{f_{\cS_i}g_{\cS_i}}(\cc_{\cS_i},\mm_{\cS_i}))>0
   \ \text{for all}\ i\in\I\\
   \zz_{\cS_i}\in\fC_{f_{\cS_i}g_{\cS_i}}(\cc_{\cS_i},\mm_{\cS_i})
   \ \text{for all}\ i\in\I\\
   g_s(\hzz_{j,s})\neq \mm_s
   \ \text{for some}\ j\in\J, s\in\D_j
 }}
 \prod_{j\in\J}
 \mu_{\hZ_{\D_j}|Y_j}(\hzz_{\D_j}|\yy_j)
 \mu_{Y_j|X_{\I}}(\yy_j|\xx_{\I})
 \prod_{i\in\I}
 \frac{
  \mu_{X_i|Z_{\cS_i}}(\xx_i|\zz_{\cS_i})
  \mu_{Z_{\cS_i}}(\zz_{\cS_i})
 }{
  \mu_{Z_{\cS_i}}(\fC_{f_{\cS_i}g_{\cS_i}}(\cc_{\cS_i},\mm_{\cS_i}))
  \prod_{s\in\cS_i}|\im\G_s|
 }
 \notag
 \\
 &=
 \sum_{\substack{
   \mm_{\cS}:\\
   \mu_{Z_{\cS}}(\fC_{f_{\cS}g_{\cS}}(\cc_{\cS},\mm_{\cS}))=0
 }}
 \prod_{i\in\I}
 \frac1{\prod_{s\in\cS_i}|\im\G_s|}
 +
 \sum_{\substack{
   \mm_{\cS},\zz_{\cS},\yy_{\J},\hzz_{\D_{\J}}:\\
   \mu_{Z_{\cS}}(\fC_{f_{\cS}g_{\cS}}(\cc_{\cS},\mm_{\cS}))>0\\
   \zz_{\cS}\in\fC_{f_{\cS}g_{\cS}}(\cc_{\cS},\mm_{\cS})\\
   g_s(\hzz_{j,s})\neq \mm_s
   \ \text{for some}\ j\in\J, s\in\D_j
 }}
 \frac{\mu_{Z_{\cS}Y_{\J}\hZ_{\D_{\J}}}(\zz_{\cS},\yy_{\J},\hzz_{\D_{\J}})}
 {
  \mu_{Z_{\cS}}(\fC_{f_{\cS}g_{\cS}}(\cc_{\cS},\mm_{\cS}))
  \prod_{s\in\cS}|\im\G_s|
 }
 \notag
 \\
 &\leq
 \sum_{\substack{
   \mm_{\cS}:\\
   \mu_{Z_{\cS}}(\fC_{f_{\cS}g_{\cS}}(\cc_{\cS},\mm_{\cS}))=0
 }}
 \prod_{i\in\I}
 \frac1{\prod_{s\in\cS_i}|\im\G_s|}
 +
 \sum_{\substack{
   \mm_{\cS},\zz_{\cS},\yy_{\J},\hzz_{\D_{\J}}:\\
   \mu_{Z_{\cS}}(\fC_{f_{\cS}g_{\cS}}(\cc_{\cS},\mm_{\cS}))>0\\
   \zz_{\cS}\in\fC_{f_{\cS}g_{\cS}}(\cc_{\cS},\mm_{\cS})\\
   g_s(\hzz_{j,s})\neq \mm_s
   \ \text{for some}\ j\in\J, s\in\D_j
 }}
 \frac{\mu_{Z_{\cS}Y_{\J}\hZ_{\D_{\J}}}(\zz_{\cS},\yy_{\J},\hzz_{\D_{\J}})}
 {\mu_{Z_{\cS}}(\fC_{f_{\cS}}(\cc_{\cS}))}
 \notag
 \\*
 &\quad
 +
 \sum_{\substack{
   \mm_{\cS},\zz_{\cS},\yy_{\J},\hzz_{\D_{\J}}:\\
   \mu_{Z_{\cS}}(\fC_{f_{\cS}g_{\cS}}(\cc_{\cS},\mm_{\cS}))>0\\
   \zz_{\cS}\in\fC_{f_{\cS}g_{\cS}}(\cc_{\cS},\mm_{\cS})\\
   g_s(\hzz_{j,s})\neq \mm_s
   \ \text{for some}\ j\in\J, s\in\D_j
 }}
 \mu_{Z_{\cS}Y_{\J}\hZ_{\D_{\J}}}(\zz_{\cS},\yy_{\J},\hzz_{\D_{\J}})
 \lrbar{
  \frac1
  {
   \mu_{Z_{\cS}}(\fC_{f_{\cS}g_{\cS}}(\cc_{\cS},\mm_{\cS}))
   \prod_{s\in\cS}|\im\G_s|
  }
  -\frac1
  {\mu_{Z_{\cS}}(\fC_{f_{\cS}}(\cc_{\cS}))}
 }
 \notag
 \\*
 &\leq
 \sum_{\substack{
   \mm_{\cS}:\\
   \mu_{Z_{\cS}}(\fC_{f_{\cS}g_{\cS}}(\cc_{\cS},\mm_{\cS}))=0\\
 }}
 \prod_{i\in\I}
 \frac1{\prod_{s\in\cS_i}|\im\G_s|}
 +
 \sum_{\substack{
   \mm_{\cS},\zz_{\cS},\hzz_{\D_{\J}}:\\
   \zz_{\cS}\in\fC_{f_{\cS}g_{\cS}}(\cc_{\cS},\mm_{\cS})\\
   \hzz_{\D_j}\neq \zz_{\D_j}
   \ \text{for some}\ j\in\J
 }}
 \frac{\mu_{Z_{\cS}\hZ_{\D_{\J}}}(\zz_{\cS},\hzz_{\D_{\J}})}
 {\mu_{Z_{\cS}}(\fC_{f_{\cS}}(\cc_{\cS}))}
 \notag
 \\*
 &\quad
 +
 \sum_{\substack{
   \mm_{\cS}:\\
   \mu_{Z_{\cS}}(\fC_{f_{\cS}g_{\cS}}(\cc_{\cS},\mm_{\cS}))>0
 }}
 \mu_{Z_{\cS}}(\fC_{f_{\cS}g_{\cS}}(\cc_{\cS},\mm_{\cS}))
 \lrbar{
  \frac1
  {
   \mu_{Z_{\cS}}(\fC_{f_{\cS}g_{\cS}}(\cc_{\cS},\mm_{\cS}))
   \prod_{s\in\cS}|\im\G_s|
  }
  -
  \frac1
  {\mu_{Z_{\cS}}(\fC_{f_{\cS}}(\cc_{\cS}))}
 }
 \notag
 \\*
 &=
 \sum_{\substack{
   \mm_{\cS},\zz_{\cS},\hzz_{\D_{\J}}:\\
   \zz_{\cS}\in\fC_{f_{\cS}g_{\cS}}(\cc_{\cS},\mm_{\cS})\\
   \hzz_{\D_j}\neq \zz_{\D_j}
   \ \text{for some}\ j\in\J
 }}
 \frac{\mu_{Z_{\cS}\hZ_{\D_{\J}}}(\zz_{\cS},\hzz_{\D_{\J}})}
 {\mu_{Z_{\cS}}(\fC_{f_{\cS}}(\cc_{\cS}))}
 +
 \sum_{\mm_{\cS}}
 \lrbar{
  \frac1
  {\prod_{s\in\cS}|\im\G_s|}
  -
  \frac{\mu_{Z_{\cS}}(\fC_{f_{\cS}g_{\cS}}(\cc_{\cS},\mm_{\cS}))}
  {\mu_{Z_{\cS}}(\fC_{f_{\cS}}(\cc_{\cS}))}
 }
 \label{eq:proof-channel-error}
\end{align}
\hrulefill
\vspace*{4pt}
\end{figure*}
\subsection{Proof of Theorem~\ref{lem:channel-encoder}}
\label{sec:proof-channel}

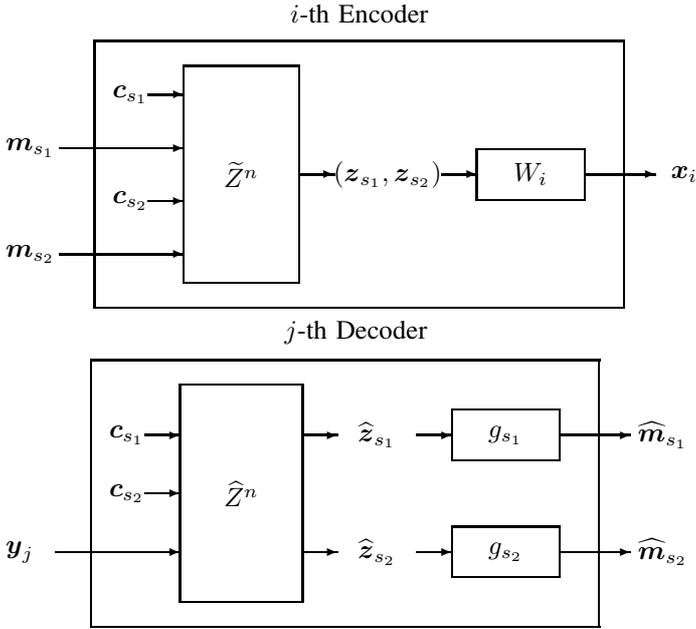
\begin{figure}[t]
\begin{center}
 \unitlength 0.47mm
 \begin{picture}(186,90)(0,6)
  \put(94,89){\makebox(0,0){$i$-th Encoder}}
  \put(20,6){\framebox(148,75){}}
  \put(30,66){\makebox(0,0){$\cc_{s_1}$}}
  \put(35,66){\vector(1,0){10}}
  \put(2,51){\makebox(0,0){$\mm_{s_1}$}}
  \put(10,51){\vector(1,0){35}}
  \put(30,36){\makebox(0,0){$\cc_{s_2}$}}
  \put(35,36){\vector(1,0){10}}
  \put(2,21){\makebox(0,0){$\mm_{s_2}$}}
  \put(10,21){\vector(1,0){35}}
  \put(45,13){\framebox(32,61){$\tZ^n$}}
  \put(77,43.5){\vector(1,0){10}}
  \put(102,43.5){\makebox(0,0){$(\zz_{s_1},\zz_{s_2})$}}
  \put(117,43.5){\vector(1,0){10}}
  \put(127,36.5){\framebox(30,14){$W_i$}}
  \put(157,43.5){\vector(1,0){20}}
  \put(185,43.5){\makebox(0,0){$\xx_i$}}
 \end{picture}
 \\
 \begin{picture}(188,90)(0,6)
  \put(94,89){\makebox(0,0){$j$-th Decoder}}
  \put(30,60){\makebox(0,0){$\cc_{s_1}$}}
  \put(35,60){\vector(1,0){10}}
  \put(30,43.5){\makebox(0,0){$\cc_{s_2}$}}
  \put(35,43.5){\vector(1,0){10}}
  \put(0,27){\makebox(0,0){$\yy_j$}}
  \put(10,27){\vector(1,0){35}}
  \put(45,13){\framebox(34,61){$\hZ^n$}}
  \put(79,60){\vector(1,0){10}}
  \put(100,60){\makebox(0,0){$\hzz_{s_1}$}}
  \put(111,60){\vector(1,0){10}}
  \put(121,53){\framebox(30,14){$g_{s_1}$}}
  \put(151,60){\vector(1,0){20}}
  \put(180,60){\makebox(0,0){$\hmm_{s_1}$}}
  \put(79,27){\vector(1,0){10}}
  \put(100,27){\makebox(0,0){$\hzz_{s_2}$}}
  \put(111,27){\vector(1,0){10}}
  \put(121,20){\framebox(30,14){$g_{s_2}$}}
  \put(151,27){\vector(1,0){20}}
  \put(180,27){\makebox(0,0){$\hmm_{s_2}$}}
  \put(20,6){\framebox(142,75){}}
 \end{picture}
\end{center}
\caption{Construction of Channel Code. For simplicity, $\cS_i=\{s_1,s_2\}$ and
 $\D_j=\{s_1,s_2\}$ are assumed.}
\label{fig:channel-code}
\end{figure}

Before the proof of the theorem, in Fig.~\ref{fig:channel-code} we
illustrate the code construction for a simple case.

Here, we prove the theorem.
For a given $\{(r_s,R_s)\}_{s\in\cS}$ that sarisfies
(\ref{eq:channel-encoder})
for all $(i,\cS_i')$ satisfying $i\in\I$ and $\emptyset\neq\cS_i'\subset\cS_i$,
assume that $(\bcG_s,\bp_{G_s})$
have a balanced-coloring property for every $s\in\cS$,
where $r_s=\log(|\im f_s|)/n$ and $R_s=\log(|\im\G_s|)/n$.

In the following,
we omit the dependence of
$Z$, $C$, $Y$, and $\hZ$ on $n$.
Let
\begin{equation*}
 \fC_{f_{\cS_i}g_{\cS_i}}(\cc_{\cS_i},\mm_{\cS_i})
 \equiv
 \fC_{f_{\cS_i}}(\cc_{\cS_i})\cap\fC_{g_{\cS_i}}(\mm_{\cS_i}).
\end{equation*}
We use the fact without notice that
$\{\fC_{f_{\cS}g_{\cS}}(\cc_{\cS},\mm_{\cS}))\}_{\cc_{\cS},\mm_{\cS}}$
is a partition of $\Z_{\cS}^n$,
and $\{\cS_i\}_{i\in\I}$ is a partition of $\cS$,

The error probability 
$\Error(f_{\cS},g_{\cS},\cc_{\cS})$ is evaluated as
(\ref{eq:proof-channel-error}),
which appears on the top of the next page,
where the first and the second terms
on the right hand side of (\ref{eq:proof-channel-error})
correspond to the encoding error 
and the decoding error probabilities, respectively.

The expectation of the first term
on the right hand side of (\ref{eq:proof-channel-error})
is evaluated as
\begin{align}
&
E_{G_{\cS}C_{\cS}}\lrB{
 \sum_{\substack{
   \mm_{\cS},\zz_{\cS},\hzz_{\D_{\J}}:\\
   \zz_{\cS}\in\fC_{f_{\cS}G_{\cS}}(C_{\cS},\mm_{\cS})\\
   \hzz_{\D_j}\neq \zz_{\D_j}
   \ \text{for some}\ j\in\J
 }}
 \frac{\mu_{Z_{\cS}\hZ_{\D_{\J}}}(\zz_{\cS},\hzz_{\D_{\J}})}
 {\mu_{Z_{\cS}}(\fC_{f_{\cS}}(C_{\cS}))}
}
\notag
\\
&=
E_{G_{\cS}}\lrB{
 \sum_{\substack{
   \cc_{\cS},\mm_{\cS},\zz_{\cS},\hzz_{\D_{\J}}:\\
   \zz_{\cS}\in\fC_{f_{\cS}G_{\cS}}(\cc_{\cS},\mm_{\cS})\\
   \hzz_{\D_j}\neq \zz_{\D_j}
   \ \text{for some}\ j\in\J
 }}
 \mu_{Z_{\cS}\hZ_{\D_{\J}}}(\zz_{\cS},\hzz_{\D_{\J}})
}
\notag
\\
&=
\sum_{\substack{
  \zz_{\cS},\hzz_{\D_{\J}}:\\
  \hzz_{\D_j}\neq \zz_{\D_j}
  \ \text{for some}\ j\in\J
}}
\mu_{Z_{\cS}\hZ_{\D_{\J}}}(\zz_{\cS},\hzz_{\D_{\J}})
\notag
\\
&=
\Error(f_{\cS}),
\label{eq:proof-channel-source}
\end{align}
where the first equality comes from
the fact that $C_{\cS}$ is generated at random
subject to the distribution
$\{\mu_{Z_{\cS}}(\fC_{f_{\cS}}(\cc_{\cS}))\}_{\cc_{\cS}}$,
and the last equality comes from the definition of $\Error(f_{\cS})$.

Let $\uT_{Z_{\cS}}$ be defined as
\begin{align*}
\uT_{Z_{\cS}}
\equiv
\lrb{
 \zz_{\cS}:
 \begin{aligned}
  &\frac 1n\log\frac1{\mu_{Z_{\cS'}}(\zz_{\cS'})}\geq \uH(\ZZ_{\cS'})-\e
  \\
  &\text{for all}\ \cS'\subset\cS
 \end{aligned}
}.
\end{align*}
Then the expectation of the second term
on the right hand side of (\ref{eq:proof-channel-error})
is evaluated as
\begin{figure*}
\normalsize
\begin{align}
&
E_{G_{\cS}C_{\cS}}\lrB{
 \sum_{\mm_{\cS}}
 \lrbar{
  \frac{\mu_{Z_{\cS}}(\fC_{f_{\cS}G_{\cS}}(C_{\cS},\mm_{\cS}))}
  {\mu_{Z_{\cS}}(\fC_{f_{\cS}}(C_{\cS}))}
  -
  \frac1{\prod_{s\in\cS}|\im\G_s|}
 }
}
\notag
\\*
&=
E_{G_{\cS}}\lrB{
 \sum_{\mm_{\cS},\cc_{\cS}}
 \lrbar{
  \mu_{Z_{\cS}}(
   \fC_{f_{\cS}}(\cc_{\cS})\cap\fC_{G_{\cS}}(\mm_{\cS}))
  -
  \frac
  {\mu_{Z_{\cS}}(\fC_{f_{\cS}}(\cc_{\cS}))}
  {\prod_{s\in\cS}|\im\G_s|}
 }
}
\notag
\\*
&\leq
E_{G_{\cS}}\lrB{
 \sum_{\mm_{\cS},\cc_{\cS}}
 \lrbar{
  \mu_{Z_{\cS}}(
   \uT_{Z_{\cS}}\cap\fC_{f_{\cS}}(\cc_{\cS})\cap\fC_{G_{\cS}}(\mm_{\cS})
  )
  -
  \frac{\mu_{Z_{\cS}}(\uT_{Z_{\cS}}\cap\fC_{f_{\cS}}(\cc_{\cS}))}
  {\prod_{s\in\cS}|\im\G_s|}
 }
}
\notag
\\*
&\quad
+
E_{G_{\cS}}\lrB{
 \sum_{\mm_{\cS},\cc_{\cS}}
 \left|
  \mu_{Z_{\cS}}(
   \uT_{Z_{\cS}}^c\cap\fC_{f_{\cS}}(\cc_{\cS})\cap\fC_{G_{\cS}}(\mm_{\cS})
  )
  -
  \sum_{\mm_{\cS},\cc_{\cS}}
  \frac{\mu_{Z_{\cS}}(\uT_{Z_{\cS}}^c\cap\fC_{f_{\cS}}(\cc_{\cS}))}
  {\prod_{s\in\cS}|\im\G_s|}
 \right|
}
\notag
\\*
&\leq
E_{G_{\cS}}\lrB{
 \sum_{\mm_{\cS},\cc_{\cS}}
 \lrbar{
  \mu_{Z_{\cS}}(
   \uT_{Z_{\cS}}\cap\fC_{f_{\cS}}(\cc_{\cS})\cap\fC_{G_{\cS}}(\mm_{\cS})
  )
  -
  \frac{\mu_{Z_{\cS}}(
    \uT_{Z_{\cS}}\cap\fC_{f_{\cS}}(\cc_{\cS})
  )}
  {\prod_{s\in\cS}|\im\G_s|}
 }
}
\notag
\\*
&\quad
+
E_{G_{\cS}}\lrB{
 \sum_{\mm_{\cS},\cc_{\cS}}
 \mu_{Z_{\cS}}(
  \uT_{Z_{\cS}}^c\cap\fC_{f_{\cS}}(\cc_{\cS})\cap\fC_{G_{\cS}}(\mm_{\cS})
 )
}
+
E_{G_{\cS}}\lrB{
 \sum_{\mm_{\cS},\cc_{\cS}}
 \frac{\mu_{Z_{\cS}}(\uT_{Z_{\cS}}^c\cap\fC_{f_{\cS}}(\cc_{\cS}))}
 {\prod_{s\in\cS}|\im\G_s|}
}
\notag
\\
&=
\sum_{\cc_{\cS}}
\mu_{Z_{\cS}}(\uT_{Z_{\cS}}\cap\fC_{f_{\cS}}(\cc_{\cS}))
E_{G_{\cS}}\lrB{
 \sum_{\mm_{\cS}}
 \lrbar{
  \frac{\mu_{Z_{\cS}}(
    \uT_{Z_{\cS}}\cap\fC_{f_{\cS}}(\cc_{\cS})\cap\fC_{G_{\cS}}(\mm_{\cS})
  )}
  {\mu_{Z_{\cS}}(\uT_{Z_{\cS}}\cap\fC_{f_{\cS}}(\cc_{\cS}))}
  -
  \frac1
  {\prod_{s\in\cS}|\im\G_s|}
 }
}
+
2\mu_{Z_{\cS}}(\uT_{Z_{\cS}}^c)
\notag
\\
&\leq
\sum_{\cc_{\cS}}
\mu_{Z_{\cS}}(\uT_{Z_{\cS}}\cap\fC_{f_{\cS}}(\cc_{\cS}))
\sqrt{
 \alpha_{G_{\cS}}-1
 +\frac{\sum_{\substack{
    \cS'\subset\cS:
    \cS'\neq\emptyset
  }}
  \alpha_{G_{\cS'^c}}
  [\beta_{G_{\cS'}}+1]\lrB{\prod_{s\in\cS'}|\im\G_s|}
  2^{-n[\uH(\ZZ_{\cS'})-\e]}
  \mu_{Z_{\cS'^c}}(\fC_{f_{\cS'^c}}(\cc_{\cS'^c}))}
 {\mu_{Z_{\cS}}(\uT_{Z_{\cS}}\cap\fC_{f_{\cS}}(\cc_{\cS}))}
}
\notag
\\*
&\quad
+
2\mu_{Z_{\cS}}(\uT_{Z_{\cS}}^c)
\notag
\\
&\leq
\mu_{Z_{\cS}}(\uT_{Z_{\cS}})
\sqrt{
 \alpha_{G_{\cS}}-1
 +
 \frac{\sum_{\cc_{\cS}}\sum_{\substack{
    \cS'\subset\cS:
    \cS'\neq\emptyset
  }}
  \alpha_{G_{\cS'^c}}
  [\beta_{G_{\cS'}}+1]\lrB{\prod_{s\in\cS'}|\im\G_s|}
  2^{-n[\uH(\ZZ_{\cS'})-\e]}
  \mu_{Z_{\cS'^c}}(\fC_{f_{\cS'^c}}(\cc_{\cS'^c}))}
 {\mu_{Z_{\cS}}(\uT_{Z_{\cS}})}
}
\notag
\\*
&\quad
+2\mu_{Z_{\cS}}(\uT_{Z_{\cS}}^c)
\notag
\\
&\leq
\sqrt{
 \alpha_{G_{\cS}}-1
 +\sum_{\substack{
   \cS'\subset\cS:
   \cS'\neq\emptyset
 }}
 \alpha_{G_{\cS'^c}}
 [\beta_{G_{\cS'}}+1]\lrB{\prod_{s\in\cS'}|\im f_s||\im\G_s|}
 2^{-n[\uH(\ZZ_{\cS'})-\e]}
}
+
2\mu_{Z_{\cS}}(\uT_{Z_{\cS}}^c)
\label{eq:proof-channel-bcp}
\end{align}
\hrulefill
\vspace*{4pt}
\end{figure*}
(\ref{eq:proof-channel-bcp}), which appears in the top of the next page.
The third inequality comes from
Lemma~\ref{lem:mBCP} by letting
\begin{align*}
\T
&\equiv
\uT_{Z_{\cS}}\cap\fC_{f_{\cS}}(\cc_{\cS})
\\
Q
&\equiv\mu_{Z_{\cS}}
\end{align*}
and using the relations
\begin{align}
\T_{\cS'}
&\subset
\lrb{
 \zz_{\cS'}:
 \frac 1n\log\frac1{\mu_{Z_{\cS'}}(\zz_{\cS'})}\geq \uH(\ZZ_{\cS'})-\e
}
\notag
\\
\T_{\cS'^c|\cS'}(\zz_{\cS'})
&\subset
\fC_{f_{\cS'^c}}(\cc_{\cS'^c})
\notag
\end{align}
as
\begin{align}
\bQ_{\cS'^c}
&=
\max_{\zz_{\cS'}\in\T_{\cS'}}
\sum_{\zz_{\cS'^c}\in\T_{\cS'^c|\cS'}(\zz_{\cS'})}
\mu_{Z_{\cS'}}(\zz_{\cS'})
\mu_{Z_{\cS'^c}}(\zz_{\cS'^c})
\notag
\\
&\leq
\lrB{\max_{\zz_{\cS'}\in\T_{\cS'}}\mu_{Z_{\cS'}}(\zz_{\cS'})}
\sum_{\zz_{\cS'^c}\in\fC_{A_{\cS'^c}}(\cc_{\cS'^c})}
\mu_{Z_{\cS'^c}}(\zz_{\cS'^c})
\notag
\\
&\leq
2^{-n[\uH(\ZZ_{\cS'})-\e]}\mu_{Z_{\cS'^c}}(\fC_{f_{\cS'^c}}(\cc_{\cS'^c})).
\end{align}
The fourth inequality comes from the Jensen inequality.
The last equality comes from the fact that
\begin{equation*}
\sum_{\cc_{\cS}}\mu_{Z_{\cS'^c}}(\fC_{f_{\cS'^c}}(\cc_{\cS'^c}))
=\prod_{s\in\cS'}|\im f_s|
\end{equation*}
and $\mu_{Z_{\cS}}(\uT_{Z_{\cS}})\leq 1$.

From (\ref{eq:proof-channel-source}) and (\ref{eq:proof-channel-bcp}),
\begin{align}
&
E_{G_{\cS}C_{\cS}}\lrB{\Error(f_{\cS},G_{\cS},C_{\cS})}
\notag
\\*
&\leq
\Error(f_{\cS})
\notag
\\*
&\quad
+
\sqrt{\textstyle
 \alpha_{G_{\cS}}-1
 +
 \sum_{\substack{
   \cS'\subset\cS:
   \cS'\neq\emptyset
 }}
 \alpha_{G_{\cS'^c}}
 \lrB{\beta_{G_{\cS'}}+1}
 2^{-n\gamma}
}
\notag
\\*
&\quad
+2\mu_{Z_{\cS}}(\uT_{Z_{\cS}}^c),
\label{eq:proof-channel-exp-error}
\end{align}
where
\begin{equation*}
\gamma \equiv \uH(\ZZ_{\cS'})-\sum_{s\in\cS'}\lrB{R_s+r_s}-\e.
\end{equation*}

Finally, let us assume that
$\{(r_s,R_s)\}_{s\in\cS}$
satisfies (\ref{eq:RIT-rR-disjoint})
for all $(i,\cS')$ satisfying 
$i\in\I$ and $\emptyset\neq\cS'\subset\cS_i$.
Then we have
\begin{align}
\sum_{s\in\cS'}\lrB{R_s+r_s}
&=
\sum_{i\in\I}\sum_{s\in\cS'\cap\cS_i}\lrB{R_s+r_s}
\notag
\\
&<
\sum_{i\in\I}\sum_{s\in\cS'\cap\cS_i}\uH(\ZZ_{\cS'\cap\cS_i})
\notag
\\
&\leq
\uH(\ZZ_{\cS'}),
\end{align} 
where
the last inequality comes from
(\ref{eq:pliminf-lower}) and the fact that
$\{Z_s\}_{s\in\cS_i}$ and $\{Z_s\}_{s\in\cS_{i'}}$
are mutually independent when $i\neq i'$.
Then, by letting $\e\to0$,
$\alpha_{G_{\cS'}}\to1$, $\log(1+\beta_{G_{\cS'}})/n\to0$
and
$\mu_{Z_{\D_j}Y_j}(\oT_{Z_{\D_j}|Y_j}^c)\to0$,
we have the fact that
for all $\delta>0$ and sufficiently large $n$
there are $\{g_s\}_{s\in\cS}$ and $\{\cc_s\}_{s\in\cS}$
such that $\Error(f_{\cS},g_{\cS},\cc_{\cS})\leq\Error(f_{\cS})+\delta$.
\hfill\IEEEQED


\begin{thebibliography}{99}
\addcontentsline{toc}{chapter}{Bibliography}
\bibitem{AC98}
R.\ Ahlswede and I.\ Csisz\'{a}r,
``Common randomness in information theory and cryptography
--- Part II: CR capacity,''
{\it IEEE Trans.\ Inform.\ Theory},
vol.\ IT-44, 
pp.\ 225--240, Jan.\ 1998.
\bibitem{C75}
T.\ M.\ Cover,
``A proof of the data compression theorem of Slepian and Wolf for ergodic sources,''
{\it IEEE Trans. Inform Theory},
vol.~IT-21, no.~2, pp.~226--228, Mar.\ 1975.
\bibitem{CSI82}
I.\ Csisz\'{a}r,
``Linear codes for sources and source networks:
Error exponents, universal coding,''
{\it IEEE Trans.\ Inform.\ Theory},
vol.\ IT-28, no.\ 4, pp.\ 585--592, Jul.\ 1982.
\bibitem{CK11}
I.\ Csisz\'{a}r and J.\ K\"{o}rner,
{\it Information Theory: Coding Theorems for Discrete Memoryless Systems
2nd Ed.},
Cambridge University Press, 2011.
\bibitem{CN03}
I.\ Csisz\'{a}r and P.\ Narayan,
``Secret key capacity for multiple terminals,''
{\it IEEE Trans.\ Inform.\ Theory},
vol.\ IT-50,  no.\ 12,
pp.\  3047--3061, Dec. 2004.
\bibitem{CW}
J.\ L.\ Carter and M.\ N.\ Wegman,
``Universal classes of hash functions,''
{\it J.\ Comput.\ Syst.\ Sci.}, vol.\ 18, pp.\ 143--154, 1979.
\bibitem{EK11}
A.\ El Gamal and Y.H. Kim,
{\it Network Information Theory},
Cambridge University Press, 2011.
\bibitem{GP80}
S.\ I.\ Gel'fand and M.\ S.\ Pinsker,
``Capacity of a broadcast channel with one deterministic component,''
{\em Probl. Inf. Transm.}, vol.\ 16, no.\ 1, pp.\ 17--25, Jan.--Mar. 1980.
\bibitem{H79}
T.S.\ Han,
``The capacity region of general multiple-access channel with certain
correlated sources,''
{\it Inform. Contr.}, vol.\ 40, pp.\ 37--60, 1979.
\bibitem{H98}
T.S.\ Han,
``An information-spectrum approach to capacity
theorems for the general multiple-access channel,''
{\it IEEE Trans.\ Inform.\ Theory}, vol.\ IT-44, pp.\ 2773--2795, Jan.\ 1998.
%
\bibitem{HAN}
T.S.\ Han,
{\it Information-Spectrum Methods in Information Theory},
Springer, 2003.
\bibitem{IZ89}
R.\ Impagliazzo and D.\ Zuckerman,
``How to recycle random bits,''
{\it 30th IEEE Symp. Fund. Computer Sci.},
Oct.~30--Nov.~1, 1989, pp.~248--253.
\bibitem{IO05}
K.\ Iwata and Y.\ Oohama,
``Information-spectrum characterization of broadcast channel with
general source,''
{\it IEICE Trans.\ Fundamentals}, vol.~E88-A, No.~10,
pp.~2808--2818, Oct.\ 2005.
\bibitem{LKP11}
Y. Liang, G. Kramer, and H. V. Poor,
``On the equivalence of two achievable regions for the broadcast
channel,''
{\it IEEE Trans. Inf. Theory}, vol.\ 57, no.\ 1, pp.\ 95--100,
Jan. 2011.
\bibitem{M79}
K.\ Marton,
``A coding theorem for the discrete memoryless broadcast channel,''
{\it IEEE Trans.\ Inform.\ Theory},
vol.\ IT-25, no.\ 3, pp. 306-311, May 1979.
\bibitem{MK95}
S.\ Miyake and F.\ Kanaya,
``Coding theorems on correlated general sources,''
{\it IEICE Trans.\ Fundamentals}, vol.\ E78-A, No.\ 9,
pp.\ 1063--1070, Sept.\ 1995.
\bibitem{ITW13}
J.\ Muramatsu,
``Equivalence between inner regions for broadcast channel coding,''
{\it Proc. 2013 IEEE Information Theory Workshop},
Seville, Spain, Sep.\ 9--13, 2013, pp.\ 164--168.
\bibitem{CRNG}
J.\ Muramatsu,
``Channel coding and lossy source coding using a generator
of constrained random numbers,''
{\it IEEE Trans.\ Inform.\ Theory},
vol.~IT-60, no.~5, pp.~2667--2686, May 2014.
\bibitem{CRNGVLOSSY}
J.\ Muramatsu,
``Variable-length lossy source code using a constrained-random-number
generator,''
{\it IEEE Trans.\ Inform.\ Theory},
vol.~IT-61, no.~6, pp.~3574--3592, Jun.\ 2015.
\bibitem{HASH}
J.\ Muramatsu and S.\ Miyake,
``Hash property and coding theorems for sparse matrices and
maximal-likelihood coding,''
{\it IEEE Trans.\ Inform.\ Theory}, vol.\ IT-56, no. 5, pp.\ 2143--2167,
May 2010.
Corrections: vol.\ IT-56, no.\ 9, p.\ 4762, Sep.\ 2010,
vol.\ IT-59, no.\ 10, pp.\ 6952--6953, Oct.\ 2013.
\bibitem{HASH-BC}
J.\ Muramatsu and S.\ Miyake,
``Construction of
Slepian-Wolf source code and broadcast channel code
based on hash property,''
available at {\tt arXiv:1006.5271[cs.IT]}, 2010.
\bibitem{ISIT2010}
J.\ Muramatsu and S.\ Miyake,
``Construction of broadcast channel code
based on hash property,''
{\it Proc.\ 2010 IEEE Int.\ Symp.\ Inform.\ Theory},
Austin, U.S.A.,
June 13--18,
pp.\ 575--579, 2010.
Extended version is available at {\tt arXiv:1006.5271[cs.IT]}, 2010.
\bibitem{SW2CC}
J.\ Muramatsu and S.\ Miyake,
``Construction of a channel code from an arbitrary source code with
decoder side information,''
{\it Proc. Int. Symp. on Inform. Theory and Its
Applicat.}, Monterey, USA, Oct.~30--Nov.~2, 2016, pp. 176--180.
Extended version is available at {\tt arXiv:1601.05879[cs.IT]}.
\bibitem{SDECODING}
J.\ Muramatsu and S.\ Miyake,
``On the error probability of stochastic decision and stochastic
decoding,''
{\it Proc.\ 2017 IEEE Int.\ Symp.\ Inform.\ Theory},
Aachen, Germany, Jun.\ 25--30, 2017, pp.\ 1643--1647.
Extended version is available at
{\tt arXiv:1701.04950[cs.IT]}.
\bibitem{SWLDPC}
J.\ Muramatsu, T.\ Uyematsu, and T.\ Wadayama,
``Low density parity check matrices for coding of correlated sources,''
{\it IEEE Trans.\ Inform.\ Theory}, vol.\ IT-51, no.\ 10,
pp.\ 3645--3653, Oct.\ 2005.
\bibitem{SW73}
D.\ Slepian and J.\ K.\ Wolf,
``Noiseless coding of correlated information sources,''
{\it IEEE Trans.\ Inform.\ Theory}, vol.\ IT-19, no.\ 4,
pp.\ 471--480, Jul.\ 1973.
\bibitem{SW73MAC}
D.\ Slepian and J.\ K.\ Wolf,
``A coding theorem for multiple access channels with correlated sources,''
{\it Bell Syst.\ Tech.\ J.}, vol.\ 52, no.\ 7,
pp.\ 1037--1076, Sep.\ 1973.
\bibitem{SV06}
A.\ Somekh-Baruch and S.\ Verd\'u,
``General relayless networks: representation of the capacity region,''
{\it Proc.\ 2006 IEEE Int.\ Symp.\ Inform.\ Theory},
Seattle, USA, 6--12 July, 2006, pp.\ 2408--2412.
\end{thebibliography}
\end{document}